\documentclass[aps,prd,reprint,twocolumn,superscriptaddress,longbibliography,nofootinbib,floatfix,showpacs]{revtex4-1}
\usepackage{epsfig}
\usepackage{amsmath,amssymb,amsfonts}
\usepackage{hyperref}
\usepackage{mathrsfs}
\usepackage{bbm}
\usepackage{slashed}
\usepackage{graphicx}
\usepackage{verbatim}

\usepackage{bm}

\usepackage{bbm}

\usepackage{rotating}

\usepackage[usenames]{color}

\definecolor{darkgreen}{rgb}{0.2,0.6,0}

\newcommand{\be}{\begin{equation}}
\newcommand{\ee}{\end{equation}}
\newcommand{\bw}{\begin{widetext}}
\newcommand{\ew}{\end{widetext}}
\newcommand{\bi}{\begin{itemize}}
\newcommand{\ei}{\end{itemize}}
\newcommand{\bea}{\begin{eqnarray}}
\newcommand{\eea}{\end{eqnarray}}
\newcommand{\bracket}[2]{\bra{#1}\,#2\rangle} 
\newcommand{\bra}[1]{\langle\,#1\,|}          
\newcommand{\ket}[1]{|\,#1\,\rangle}          
\newcommand{\ud}{\mathrm{d}}

\newcommand{\LCm}{{\scriptscriptstyle -}} 
\newcommand{\LCp}{{\scriptscriptstyle +}}
\newcommand{\LCpm}{{\scriptscriptstyle \pm}}
\newcommand{\LCmp}{{\scriptscriptstyle \mp}}

\newcommand{\LCperp}{{\scriptscriptstyle \perp}}

\usepackage[T1]{fontenc} \usepackage[latin1]{inputenc}

\begin{document}

\title{Approximating higher-order nonlinear QED processes with first-order building blocks}

\author{Victor Dinu} 
\email{dinu@barutu.fizica.unibuc.ro}
\affiliation{Department of Physics, University of Bucharest, P.O.~Box MG-11, M\u agurele 077125, Romania}

\author{Greger Torgrimsson}
\email{greger.torgrimsson@uni-jena.de}
\email{g.torgrimsson@hzdr.de}
\affiliation{Theoretisch-Physikalisches Institut, Abbe Center of Photonics,
Friedrich-Schiller-Universit\"at Jena, Max-Wien-Platz 1, D-07743 Jena, Germany}
\affiliation{Helmholtz Institute Jena, Fr\"obelstieg 3, D-07743 Jena, Germany}
\affiliation{Helmholtz-Zentrum Dresden-Rossendorf, Bautzner Landstra{\ss}e 400, 01328 Dresden, Germany}

\begin{abstract}
Higher-order tree-level processes in strong laser fields, i.e. cascades, are in general extremely difficult to calculate, but in some regimes the dominant contribution comes from a sequence of first-order processes, i.e. nonlinear Compton scattering and nonlinear Breit-Wheeler pair production. At high intensity the field can be treated as locally constant, which is the basis for standard particle-in-cell codes.
However, the locally-constant-field (LCF) approximation and these particle-in-cell codes cannot be used when the intensity is only moderately high, which is a regime that is experimentally relevant.
We have shown that one can still use a sequence of first-order processes to estimate higher orders at moderate intensities provided the field is sufficiently long.
An important aspect of our new ``gluing'' approach is the role of the spin/polarization of intermediate particles, which is more nontrivial compared to the LCF regime. 
\end{abstract}
\maketitle

\section{Introduction}

Consider a single high-energy electron that collides with a high-intensity laser field. Many electrons, positrons and photons can be produced in such collisions~\cite{Bell:2008zzb,Elkina:2010up,Nerush:2010fe}. One can approximate the laser by a pulsed plane wave, and, thanks to the simplicity of the Volkov solution to the Dirac equation in such fields, one can derive compact expressions for e.g. nonlinear Compton scattering $e^\LCm\to e^\LCm+\gamma$ for arbitrary pulse shapes and parameter values. After the first photon emission the photon can decay via nonlinear Breit-Wheeler pair production~\cite{Reiss62,Nikishov:1964zza} $\gamma\to e^\LCm+e^\LCp$ or the electron can emit a second photon. The first gives one part of the nonlinear trident process~\cite{Dinu:2017uoj,Dinu:2019wdw,King:2018ibi,Mackenroth:2018smh,Acosta:2019bvh,Krajewska15,Hu:2014ooa,King:2013osa,Ilderton:2010wr,Hu:2010ye,Bamber:1999zt,Ritus:1972nf,Baier} $e^\LCm\to e^\LCm+e^\LCm+e^\LCp$, and the second gives one part of double nonlinear Compton scattering~\cite{Morozov:1975uah,Lotstedt:2009zz,Loetstedt:2009zz,Seipt:2012tn,Mackenroth:2012rb,King:2014wfa,Dinu:2018efz,Wistisen:2019pwo} $e^\LCm\to e^\LCm+\gamma+\gamma$. Since all the processes we consider here are in general nonlinear in the interaction with the laser, we will drop ``nonlinear'' from here on.
These two second order processes are significantly harder to calculate than the first-order processes. In fact, even in a plane wave, it is a challenge to calculate all contributions to the total/integrated probability e.g. for a long pulse. So, at third and higher orders one definitely needs some way to approximate the exact result. 

One regime that allows for such an approximation is the high-intensity regime. More precisely, the classical nonlinearity parameter $a_0=E/\omega$ should be large\footnote{All energies are given in units of the electron mass $m=1$, and the electron charge $e$ is absorbed into the field strength $eE\to E$.}. Exactly how large it has to be depends on the values of the other parameters in the system~\cite{Dinu:2015aci,DiPiazza:2017raw,Podszus:2018hnz,Ilderton:2019kqp}, but assuming it is large enough, then one can make an expansion in $1/a_0$. The formation length is small in this regime, so the field can be treated as approximately constant during particle production. So, in for example trident, photon emission would happen at one constant value of the field strength, the photon would propagate a macroscopic distance and then decay into a pair at another constant field strength. We call this the two-step part, and refer to the rest as one-step terms. In the production of $N$ particles we use ``$N$-step'' to refer to the corresponding cascade part.
This is the basis of particle-in-cell codes~\cite{RidgersCode,Gonoskov:2014mda,Osiris,Smilei}.    

In most of the standard PIC codes so far, whether or not to produce a particle is determined based on probabilities/rates that are summed/averaged over the spin/polarization at each step. However, it is known, see~\cite{Ritus:1972nf,Baier,King:2013osa}, how the spin/polarization of intermediate particles in trident and double Compton can be included. For constant fields this is achieved with single sums for each intermediate particle. For example, in trident one would have a single sum over two different polarization vectors of the intermediate photon. Recently, spin effects have started to be included in PIC codes~\cite{DelSorbo:2017fod,Seipt:2018adi,Seipt:2018adi,Li:2018fcz,Li:2019oxr}. See also~\cite{Ilderton:2018nws,DiPiazza:2018bfu} for further improvements on standard LCF/PIC methods.

However, if $a_0$ is not large then one cannot use this LCF approach. In this paper we provide a generalization of the LCF approach to $a_0\sim1$. For $a_0\sim1$ one can in general expect important corrections to the $N$-step. 
However, if the field has $a_0\sim1$ {\it and} a sufficiently long pulse length then the $N$-step again gives the dominant contribution, because the intermediate particles can propagate macroscopic distances, while in the corrections to the $N$-step at least some of the particles are forced to stay close. Note that what we here (with $a_0\sim1$) mean by the $N$-step is different from its LCF approximation, and its relation to the product of nonlinear Compton and Breit-Wheeler is more complicated than in the LCF regime and what is put into standard 
PIC codes. Thus, while things are simpler for a long pulse, we still have new features compared to the LCF regime.
For arbitrarily polarized fields the role of the spin/polarization of intermediate particles is significantly different from the LCF case.
Our motivation for considering long pulses with moderately high intensity comes from upcoming experiments such as LUXE~\cite{LUXEparameters,Abramowicz:2019gvx} and FACET-II~\cite{MeurenPresentation}.

This paper is organized as follows.
In Sec.~\ref{DefinitionSection} we introduce some notation and basic definitions.
In Sec.~\ref{gluing-section} we first study the relation between the dominant contribution to trident for long pulses and the incoherent product of nonlinear Compton scattering and Breit-Wheeler pair production. Then we study how this generalizes to other higher-order processes. We present two equivalent formulations of the gluing approach. 
In Sec.~\ref{saddlePointSection} we show how higher-order processes can be approximated with a saddle-point method, we show in particular how the intricate pattern in the spectrum at low energy can be understood in terms of a large number of complex saddle points.

\section{Formalism}\label{DefinitionSection}

In order to introduce notation we present the dominant contribution to the trident probability $\mathbb{P}$. The vector potential is given by $a_\LCperp(\phi)$, where $\phi=kx=\omega x^\LCp=\omega(x^0+x^3)$ and $a_\LCperp=\{a_1,a_2\}$. We also use the notation $v^\LCpm=2v_\LCmp=v^+\pm v^3$.
The initial electron has momentum $p_\mu$, the two electrons in the final state have momenta $p_{1\mu}$ and $p_{2\mu}$, and the positron has $p_{3\mu}$. Because the field only depends on one space-time coordinate, $x^\LCp$, we have a delta function $\delta^2_\LCperp(p-p_1-p_2-p_3)\delta(k[p-p_1-p_2-p_3])$, which can be used to perform e.g. the integral over the positron momentum components. 
In Sec.~\ref{gluing-section} we will show how to approximate higher-order processes by incoherently gluing together a sequence of nonlinear Compton and Breit-Wheeler steps. To do that we need the spin, polarization and longitudinal momentum dependence of these first-order building blocks. However, we can integrate the first-order building blocks over the transverse momenta ($P_\LCperp$) without losing necessary information for building higher orders. This is related to the fact that after integrating over the transverse momenta of the final state particles, the probability no longer depends on the transverse momentum of the initial particle.

So, we perform the integrals over $p_{1\LCperp}$ and $p_{2\LCperp}$ and we are left with the longitudinal momentum spectrum
\be
\mathbb{P}=\int_0^1\ud s_1\ud s_2\theta(s_3)\mathbb{P}(s) \;,
\ee
where $s_i=kp_i/kp$ is the ratio of the longitudinal momentum of particle $i$ and the initial electron. For the initial electron we use $b_0=kp$ and for the intermediate photon $q_1=1-s_1$.
In this paper we are interested in the two-step part of $\mathbb{P}$. This comes from a term that on the amplitude level has two lightfront time $x^\LCp$ integrals, so on the probability level we have 4 lightfront time integrals. We use $\phi_1$ and $\phi_2$ for the photon emission step and $\phi_3$ and $\phi_4$ for the pair production step. We find 
\be\label{P2start}
\begin{split}
	\mathbb{P}_{\rm dir}^{22}(s)=-&\frac{\alpha^2}{8\pi^2b_0^2}\int\!\ud^4\phi\frac{\theta(\theta_{42})\theta(\theta_{31}) }{q_1^2\theta_{21}\theta_{43}} 
	e^{\frac{i}{2b_0}\left[r_1\Theta_{21}+r_2\Theta_{43}\right]} \\
	&\bigg\{\frac{\kappa_{01}\kappa_{23}}{4}W_{12}W_{34}+W_{13}W_{24}+W_{14}W_{23}
	\\
	&+\left[\frac{\kappa_{01}}{2}\left(\frac{2ib_0}{r_1\theta_{21}}+1+D_1\right)-1\right] \\
	&\left[\frac{\kappa_{23}}{2}\left(\frac{2ib_0}{r_2\theta_{43}}+1+D_2\right)+1\right]-D_1D_2\bigg\} \\
	&+(s_1\leftrightarrow s_2) \;,
\end{split}
\ee
where $\ud^4\phi=\ud\phi_1...\ud\phi_4$, $\theta_{ij}:=\phi_i-\phi_j$, $r_1:=(1/s_1)-(1/s_0)$, $r_2:=(1/s_2)+(1/s_3)$,
$\kappa_{ij}=(s_i/s_j)+(s_j/s_i)$, and where $s_0=1$ is inserted for symmetry reasons. The singularities at $\theta_{21}=0$ and $\theta_{43}=0$ are regulated by $\phi_{2,4}\to\phi_{2,4}+i\epsilon$ and $\phi_{1,3}\to\phi_{1,3}-i\epsilon$ with $\epsilon>0$, or equivalent integration contours. 	 
The field enters the exponent in $\Theta_{ij}:=\theta_{ij}M^2_{ij}$ via the ``effective mass''~\cite{Kibble:1975vz} $M$, which is obtained from the lightfront-time average of the Lorentz momentum as 
\be\label{LorentzMomentumAndM2}
M^2:=\langle\pi\rangle^2 \;,
\ee
where
\be
\pi_\mu=p_\mu-a_\mu+\frac{2ap-a^2}{2kp}k_\mu \;,
\ee
and
\be
\langle F\rangle_{ij}:=\frac{1}{\theta_{ij}}\int_{\phi_j}^{\phi_i}\!\ud\phi\; F(\phi) \;.
\ee  
The prefactor depends on the field through
\be
{\bf\Delta}_{ij}:={\bf a}(\phi_i)-\langle{\bf a}\rangle_{ij} \;,
\ee
which enters via
$D_1={\bf\Delta}_{12}\!\cdot\!{\bf\Delta}_{21}$, $D_2={\bf\Delta}_{34}\!\cdot\!{\bf\Delta}_{43}$, and 
\be\label{Wijdef}
\begin{split}
	W_{ij}:=&\frac{1}{b_0}\varepsilon^{\mu\nu\rho\sigma}p_\mu k_\nu w_{i\rho} w_{j\sigma} \\
	=&w_{i1}w_{j2}-w_{i2}w_{j1}=\hat{\bf z}\!\cdot\!({\bf w}_i\!\times\!{\bf w}_j) \;,
\end{split}
\ee
where 
\be
{\bf w}_1={\bf\Delta}_{12} \quad {\bf w}_2={\bf\Delta}_{21} \quad {\bf w}_3={\bf\Delta}_{34} 
\quad {\bf w}_4={\bf\Delta}_{43}
\ee
and where $\varepsilon$ is the Levi-Civita tensor. Note that $W_{ij}=0$ for linear polarization.

The step function combination can be expressed as
\be\label{StepsForStepsSep}
\begin{split}
&\theta(\theta_{42})\theta(\theta_{31}) \\
&=\theta(\sigma_{43}-\sigma_{21})\left\{1-\theta\left(\frac{|\theta_{43}-\theta_{21}|}{2}-[\sigma_{43}-\sigma_{21}]\right)\right\} \;, 
\end{split}
\ee  
where $\sigma_{ij}=(\phi_i+\phi_j)/2$. Replacing $\theta(\theta_{42})\theta(\theta_{31})$ with $\theta(\sigma_{43}-\sigma_{21})$ gives the same result to leading order in the pulse length or in an $1/a_0$ expansion. It is natural to make this replacement because the second term in~\eqref{StepsForStepsSep} scales linearly in the volume and is therefore naturally combined with the other one-step terms. 

Note that we derived~\eqref{P2start} in~\cite{Dinu:2017uoj} without reference to the first-order processes, nonlinear Compton and Breit-Wheeler. We showed in~\cite{Dinu:2017uoj} that for linear polarization~\eqref{P2start} can be obtained from the incoherent product of these first-order processes with a single sum over the polarization of the intermediate photon. In Sec.~\ref{gluing-section} we will consider arbitrary polarization, where things are more nontrivial.

\section{Gluing approach}\label{gluing-section}

In this section we present our new gluing approach, where higher-order processes are approximated by linking together the spin and polarization dependent probabilities of nonlinear Compton scattering and Breit-Wheeler pair production. This is a generalization of the case in~\cite{Dinu:2018efz}, where we considered processes with only intermediate electrons. We expect this to give a good approximation for sufficiently long pulses and/or large $a_0$.
To treat the spin and polarization we use the following basis.
For fermions we choose
\begin{align}
\gamma^0=&
\begin{pmatrix}
0&0&1&0 \\
0&0&0&1 \\
1&0&0&0 \\
0&1&0&0
\end{pmatrix}
&
\gamma^1=&
\begin{pmatrix}
0&0&0&1 \\
0&0&1&0 \\
0&-1&0&0 \\
-1&0&0&0
\end{pmatrix}
\nonumber\\
\gamma^2=&
\begin{pmatrix}
0&0&0&-i \\
0&0&i&0 \\
0&i&0&0 \\
-i&0&0&0
\end{pmatrix}
&
\gamma^3=&
\begin{pmatrix}
0&0&1&0 \\
0&0&0&-1 \\
-1&0&0&0 \\
0&1&0&0
\end{pmatrix} 
\;
\end{align}
and (cf.~\cite{Kogut:1969xa,Brodsky:1997de})
\be
u_{\scriptscriptstyle\uparrow}=\frac{1}{\sqrt{2p_\LCm}}\begin{pmatrix}1\\0\\2p_\LCm\\-p_1-ip_2 \end{pmatrix} \qquad
u_{\scriptscriptstyle\downarrow}=\frac{1}{\sqrt{2p_\LCm}}\begin{pmatrix}p_1-ip_2\\2p_\LCm\\0\\1 \end{pmatrix} \;,
\ee
\be
v_{\scriptscriptstyle\uparrow}=\frac{1}{\sqrt{2p_\LCm}}\begin{pmatrix}1\\0\\-2p_\LCm\\p_1+ip_2 \end{pmatrix} \qquad
v_{\scriptscriptstyle\downarrow}=\frac{1}{\sqrt{2p_\LCm}}\begin{pmatrix}p_1-ip_2\\2p_\LCm\\0\\-1 \end{pmatrix} \;.
\ee
These spinors are convenient because of their simple dependence on $p_\LCperp$\footnote{Recall that $p_\LCm$ is independent on $p_\LCperp$ while $p_\LCp=(1+p_\LCperp^2)/(4p_\LCm)$, which also means that factors of $p_0=p_\LCm+p_\LCp$ could lead to more complicated expressions.}, so all integrals over $p_{1,2,\LCperp}$ are Gaussian.
A general spin state can be expressed as
\be\label{generalu}
u=\cos\left(\frac{\rho}{2}\right)u_{\scriptscriptstyle\uparrow}+\sin\left(\frac{\rho}{2}\right)e^{i\lambda}u_{\scriptscriptstyle\downarrow} \;,
\ee
\be\label{generalv}
v=\cos\left(\frac{\rho}{2}\right)v_{\scriptscriptstyle\uparrow}+\sin\left(\frac{\rho}{2}\right)e^{i\lambda}v_{\scriptscriptstyle\downarrow} \;.
\ee
We assume that $\rho$ and $\lambda$ do not depend on the momentum $p$, and then we integrate over all the transverse momentum components. It turns out that the results can be expressed neatly in terms of the following vector 
\be\label{initialSpinDirection}
\begin{split}
{\bm n}:=&\frac{1}{2}u^\dagger{\bm\Sigma}u({\bf p}=0)=\frac{1}{2}v^\dagger{\bm\Sigma}v({\bf p}=0) \\
=&\{\cos\lambda\sin\rho,\sin\lambda\sin\rho,\cos\rho\} \;,
\end{split}
\ee
where the spin matrix is given by
\be
{\bm\Sigma}=i\{\gamma^2\gamma^3,\gamma^3\gamma^1,\gamma^1\gamma^2\} \;.
\ee 
We also have
\be
u\bar{u}=\frac{1}{2}(\slashed{p}+1)(1+\gamma^5\slashed{\alpha}) \;,
\ee
where $\gamma^5=i\gamma^0\gamma^1\gamma^2\gamma^3$ and the spin 4-vector $\alpha_\mu$ is a linear combination of three basis vectors $\alpha^{(i)}p=0$ with the components of ${\bf n}$ as coefficients,
\be
\alpha_\mu=\sum_{i=1}^3 n_i\alpha^{(i)}_\mu
\ee
with (cf.~\cite{Seipt:2018adi})
\be
\begin{split}
\alpha^{(1)}_\mu&=\hat{x}_\mu-\frac{p_1}{kp}k_\mu \\
\alpha^{(2)}_\mu&=\hat{y}_\mu-\frac{p_2}{kp}k_\mu \\
\alpha^{(3)}_\mu&=p_\mu-\frac{1}{kp}k_\mu 
\end{split}
\ee
where $\hat{x}v=v_1$, $\hat{y}v=v_2$ and $\alpha^{(i)}\alpha^{(j)}=-\delta_{ij}$. Thus, ${\bf n}$ is the Stokes vector with respect to the spin basis given by $\alpha^{(i)}$. 

For a photon with momentum $l_\mu$ we choose a polarization vector with $\epsilon_\LCm=0$, $\epsilon_\LCp=l_\LCperp\epsilon_\LCperp/(2l_\LCm)$ and
\be\label{epsfromrholambda}
\epsilon_\LCperp=\left\{\cos\left(\frac{\rho}{2}\right),\sin\left(\frac{\rho}{2}\right)e^{i\lambda}\right\} \;.
\ee
We again keep $\rho$ and $\lambda$ constant while integrating over the transverse momenta.
Similar to the fermion case, we find that the polarization dependence can be expressed in terms of another three-dimensional unit vector,
\be\label{nPhotonDefinition}
\begin{split}
{\bf n}=&\epsilon_i^*\{\sigma_1,\sigma_2,\sigma_3\}_{ij}\epsilon_j \\
=&\{\cos\lambda\sin\rho,\sin\lambda\sin\rho,\cos\rho\}\;,
\end{split}
\ee
where the Pauli matrices are as usual given by
\be
{\bm\sigma}_1=\begin{pmatrix}0&1\\1&0\end{pmatrix}
\qquad
{\bm\sigma}_2=\begin{pmatrix}0&-i\\i&0\end{pmatrix}
\qquad
{\bm\sigma}_3=\begin{pmatrix}1&0\\0&-1\end{pmatrix} \;.
\ee
The meaning of this vector is of course different from the fermion case and it transforms differently under e.g. a rotation of the transverse coordinates. However, the role this vector plays in linking together the first order processes is basically the same as in the fermion case. 
${\bf n}$ is the Stokes vector with respect to the two polarizations given by $\epsilon^{(1)}_\LCperp=\{1,0\}$ and $\epsilon^{(2)}_\LCperp=\{0,1\}$. Spin and polarization effects have been studied in many papers, see for example~\cite{Fano,LippsTolhoekI,LippsTolhoekII,McMasterRevModPhys,Misaki2000,Ivanov:2004fi,Ivanov:2004vh,Cantatore:1991sq,Wistisen:2013waa,Bragin:2017yau,DelSorbo:2017fod,Seipt:2018adi,Seipt:2018adi,Li:2018fcz,Li:2019oxr}.

\subsection{Averaging approach}

With these vectors we find that the probability of nonlinear Compton scattering and Breit-Wheeler pair production can be expressed as
\be\label{PCnnn}
\begin{split}
\mathbb{P}=&\langle\mathbb{P}\rangle+{\bf n}_0\!\cdot\!{\bf P}_0+{\bf n}_1\!\cdot\!{\bf P}_1+{\bf n}_2\!\cdot\!{\bf P}_2 \\
&+{\bf n}_0\!\cdot\!{\bf P}_{01}\!\cdot\!{\bf n}_1+{\bf n}_0\!\cdot\!{\bf P}_{02}\!\cdot\!{\bf n}_2+{\bf n}_1\!\cdot\!{\bf P}_{12}\!\cdot\!{\bf n}_2 \\
&+{\bf P}_{012,ijk}{\bf n}_{0i}{\bf n}_{1j}{\bf n}_{2k}\;,
\end{split}
\ee 
with two ${\bf n}_i$'s for the fermions and one for the photon (cf. Sec.~87 in~\cite{QED-book} for ordinary Compton scattering). $\langle\mathbb{P}\rangle$ gives the spin and polarization averaged probability, ${\bf P}_i$ gives the dependence on the spin/polarization of one particle when averaging over the spin/polarization of the other two particles, ${\bf P}_{ij}$ describes the correlation between the spin/polarization of two particles, and ${\bf P}_{ijk}$ describes the correlation between the spin/polarization of all three particles.   

For Compton scattering we find that the probability that an electron with longitudinal momentum $s_0$ and spin vector ${\bf n}_0$ scatters into a state with $s_1$ and ${\bf n}_1$ by emitting a photon with momentum $q_1$ and polarization ${\bf n}_\gamma$ is given by
\be\label{PCfromRC}
\mathbb{P}^{\rm C}=:\frac{i\alpha}{8\pi b_0s_0^2}\int\frac{\ud\phi_{12}}{\theta_{21}}\exp\left\{\frac{ir}{2b_0}\Theta_{21}\right\}\mathbb{R}^{\rm C} \;,
\ee
where $\mathbb{R}^{\rm C}$ is given, with the same notation as in~\eqref{PCnnn}, by 
\be\label{CaveR}
\langle\mathbb{R}^{\rm C}\rangle=\frac{\kappa}{2}\left[\frac{2ib_0}{r\theta}+1+D_1\right]-1
\ee 
\be\label{C0R}
{\bf R}^{\rm C}_0=\frac{q_1}{s_0}\left\{{\bf 1}+\left[1+\frac{s_0}{s_1}\right]\hat{\bf k}\,{\bf X}\right\}\!\cdot\!{\bf V}
\ee
\be\label{C1R}
{\bf R}^{\rm C}_1=\frac{q_1}{s_1}\left\{{\bf 1}+\left[1+\frac{s_1}{s_0}\right]\hat{\bf k}\,{\bf X}\right\}\!\cdot\!{\bf V}
\ee
\be\label{CgammaR}
{\bf R}^{\rm C}_{\gamma,k}={\bf w}_{1}\!\cdot\!\left\{{\bf S}_k+\frac{\kappa}{2}\delta_{k2}{\bm\sigma}_2\right\}\!\cdot\!{\bf w}_2
\ee
\be\label{C01R}
\begin{split}
{\bf R}^{\rm C}_{01}=&\frac{q_1}{s_0s_1}\bigg\{s_0\hat{\bf k}\,{\bf X}-s_1{\bf X}\,\hat{\bf k}-\frac{q_1}{2}\hat{\bf k}\,\hat{\bf k}\bigg\}  \\
&+\left[\frac{2ib_0}{r\theta}+D_1\right]\left[{\bf1}_2+\frac{\kappa}{2}\hat{\bf k}\,\hat{\bf k}\right]
\end{split}
\ee
\be
\begin{split}
{\bf R}^{\rm C}_{\gamma0,k}&=\frac{q_1}{s_0s_1}\bigg\{\!-s_0{\bf S}_k\!\cdot\!{\bf V}
+s_1\delta_{k2}\bigg[{\bf X} \\
&+\left(\frac{1}{2}\left[1+\frac{s_0}{s_1}\right]\left[\frac{2ib_0}{r\theta}+D_1+1\right]-1\right)\hat{\bf k}\bigg]\bigg\}
\end{split}
\ee
\be
\begin{split}
{\bf R}^{\rm C}_{\gamma1,k}&=\frac{q_1}{s_0s_1}\bigg\{\!-s_1{\bf S}_{k}\!\cdot\!{\bf V}
+s_0\delta_{k2}\bigg[{\bf X} \\
&+\left(\frac{1}{2}\left[1+\frac{s_1}{s_0}\right]\left[\frac{2ib_0}{r\theta}+D_1+1\right]-1\right)\hat{\bf k}\bigg]\bigg\}
\end{split}
\ee
\be\label{Cgamma01R}
\begin{split}
{\bf R}^{\rm C}_{\gamma01,kij}&=\frac{q_1}{s_0s_1}\bigg\{\!s_1\hat{\bf k}\,{\bf S}_k\!\cdot\!{\bf X}-s_0{\bf S}_k\!\cdot\!{\bf X}\,\hat{\bf k}-\frac{q_1}{2}{\bf S}_k \\ 
&\hspace{1.5cm}+\delta_{k2}[s_0\hat{\bf k}\,{\bf V}-s_1{\bf V}\,\hat{\bf k}]\bigg\}_{ij} \\ 
&+{\bf w}_1\!\cdot\!\bigg\{\delta_{k2}{\bm \sigma}_2\bigg[{\bf 1}_{2,ij}+\frac{\kappa}{2}\hat{\bf k}_i\hat{\bf k}_j\bigg]
 \\
&+{\bf S}_k\bigg[\frac{\kappa}{2}{\bf 1}_{2,ij}+\hat{\bf k}_i\hat{\bf k}_j\bigg]+\frac{\tilde{\kappa}}{2}{\bm \sigma}_2\!\cdot\!{\bf S}_k\,{\bm \sigma}_{2ij}\bigg\}\!\cdot\!{\bf w}_2 \;,
\end{split}
\ee
where $r=(1/s_1)-(1/s_0)$, $\kappa=(s_0/s_1)+(s_1/s_0)$, $\tilde{\kappa}=(s_0/s_1)-(s_1/s_0)$, $\hat{\bf k}=\{0,0,1\}$,
${\bf S}_{kij}=\delta_{k1}{\bm\sigma}_{1ij}+\delta_{k3}{\bm\sigma}_{3ij}$, ${\bm\sigma}_{n,3i}={\bm\sigma}_{n,i3}=0$,
\be
{\bm1}_2=\begin{pmatrix}1&0&0\\0&1&0\\0&0&0\end{pmatrix} \;,
\ee  
and
\be\label{XVdef}
{\bf X}=\frac{1}{2}({\bf w}_2+{\bf w}_1)
\qquad
{\bf V}=\frac{1}{2}{\bm\sigma}_2\!\cdot\!({\bf w}_2-{\bf w}_1) \;.
\ee
The corresponding expressions for Compton scattering by a positron can be obtained either by 1) replacing ${\bf a}\to-{\bf a}$ and ${\bf n}\to-{\bf n}$ for the two spin vectors, or 2) by replacing $s_0\leftrightarrow-s_1$ (except in the overall factor of $1/s_0^2$ in~\eqref{PCfromRC}),  ${\bf n}_{03}\leftrightarrow{\bf n}_{13}$ and ${\bf n}_{0\LCperp}\leftrightarrow-{\bf n}_{1\LCperp}$. 

For Breit-Wheeler we find that the probability that a photon with $q_1$ and ${\bf n}_\gamma$ decays into an electron with $s_2$ and ${\bf n}_2$ and a positron with $s_3$ and ${\bf n}_3$ is given by 
\be
\mathbb{P}^{\rm BW}=:\frac{i\alpha}{8\pi b_0q_1^2}\int\frac{\ud\phi_{12}}{\theta_{21}}\exp\left\{\frac{ir}{2b_0}\Theta_{21}\right\}\mathbb{R}^{\rm BW} \;,
\ee
where
\be\label{BWaveR}
\langle\mathbb{R}^{\rm BW}\rangle=\frac{\kappa}{2}\left[\frac{2ib_0}{r\theta}+1+D_1\right]+1
\ee 
\be
{\bf R}^{\rm BW}_2=\frac{q_1}{s_2}\left\{{\bf 1}+\left[1-\frac{s_2}{s_3}\right]\hat{\bf k}\,{\bf X}\right\}\!\cdot\!{\bf V}
\ee
\be
{\bf R}^{\rm BW}_3=\frac{q_1}{s_3}\left\{{\bf 1}-\left[1-\frac{s_3}{s_2}\right]\hat{\bf k}\,{\bf X}\right\}\!\cdot\!{\bf V}
\ee
\be\label{BWgammaR}
{\bf R}^{\rm BW}_{\gamma,k}=-{\bf w}_{1}\!\cdot\!\left\{{\bf S}_k+\frac{\kappa}{2}\delta_{k2}{\bm\sigma}_2\right\}\!\cdot\!{\bf w}_2
\ee
\be
\begin{split}
{\bf R}^{\rm BW}_{23}=&\frac{q_1}{s_2s_3}\bigg\{\!-s_2\hat{\bf k}\,{\bf X}+s_3{\bf X}\,\hat{\bf k}-\frac{q_1}{2}\hat{\bf k}\,\hat{\bf k}\bigg\}  \\
&+\left[\frac{2ib_0}{r\theta}+D_1\right]\left[{\bf1}_2+\frac{\kappa}{2}\hat{\bf k}\,\hat{\bf k}\right]
\end{split}
\ee
\be
\begin{split}
{\bf R}^{\rm BW}_{\gamma2,k}&=\frac{q_1}{s_2s_3}\bigg\{\!s_2{\bf S}_k\!\cdot\!{\bf V}
-s_3\delta_{k2}\bigg[{\bf X} \\
&+\left(\frac{1}{2}\left[1-\frac{s_2}{s_3}\right]\left[\frac{2ib_0}{r\theta}+D_1+1\right]-1\right)\hat{\bf k}\bigg]\bigg\}
\end{split}
\ee
\be
\begin{split}
{\bf R}^{\rm BW}_{\gamma3,k}&=\frac{q_1}{s_2s_3}\bigg\{\!s_3{\bf S}_k\!\cdot\!{\bf V}
-s_2\delta_{k2}\bigg[{\bf X} \\
&+\left(-\frac{1}{2}\left[1-\frac{s_3}{s_2}\right]\left[\frac{2ib_0}{r\theta}+D_1+1\right]+1\right)\hat{\bf k}\bigg]\bigg\}
\end{split}
\ee
\be
\begin{split}
{\bf R}^{\rm BW}_{\gamma23,kij}&=\frac{q_1}{s_2s_3}\bigg\{\!s_3\hat{\bf k}\,{\bf S}_k\!\cdot\!{\bf X}-s_2{\bf S}_k\!\cdot\!{\bf X}\,\hat{\bf k}+\frac{q_1}{2}{\bf S}_k \\ 
&\hspace{1.5cm}+\delta_{k2}[s_2\hat{\bf k}\,{\bf V}-s_3{\bf V}\,\hat{\bf k}]\bigg\}_{ij} \\ 
&-{\bf w}_1\!\cdot\!\bigg\{\delta_{k2}{\bm \sigma}_2\bigg[{\bf 1}_{2,ij}+\frac{\kappa}{2}\hat{\bf k}_i\hat{\bf k}_j\bigg]
\\
&+{\bf S}_k\bigg[\frac{\kappa}{2}{\bf 1}_{2,ij}+\hat{\bf k}_i\hat{\bf k}_j\bigg]+\frac{\tilde{\kappa}}{2}{\bm \sigma}_2\!\cdot\!{\bf S}_k\,{\bm \sigma}_{2ij}\bigg\}\!\cdot\!{\bf w}_2 \;,
\end{split}
\ee
where $r=(1/s_2)+(1/s_3)$, $\kappa=(s_2/s_3)+(s_3/s_2)$ $\tilde{\kappa}=(s_2/s_3)-(s_3/s_2)$.
These expressions for $\mathbb{P}^{\rm BW}$ can be obtained from~\eqref{PCfromRC} to~\eqref{Cgamma01R} by replacing $s_0\to-s_3$, $s_1\to s_2$, $q_1\to-q_1$, ${\bf n}_{03}\to{\bf n}_{33}$, ${\bf n}_{0\LCperp}\to-{\bf n}_{3\LCperp}$, ${\bf n}_1\to{\bf n}_2$, ${\bf n}_{\gamma2}\to-{\bf n}_{\gamma2}$ and ${\bf n}_{\gamma1,3}\to{\bf n}_{\gamma1,3}$, and finally multiplying with an overall $-1$. The sign change for one of the components of ${\bf n}_\gamma$ can be understood as a consequence of the fact that, when changing an incoming photon to an outgoing one, one takes the complex conjugate of the polarization vector $\epsilon_\mu$, which in~\eqref{epsfromrholambda} corresponds to $\lambda\to-\lambda$, which in turn changes the sign of ${\bf n}_{\gamma2}$ in~\eqref{nPhotonDefinition}.  

The goal is now to link together these first-order terms to approximate higher-order processes for sufficiently long pulses or large $a_0$. It might seem like we have a quite large number of terms compared to the familiar LCF case, but note that the idea is that these are all the terms we need to construct the $N$-th step for any higher-order process for $a_0\gtrsim1$ and arbitrary field polarization.

We start with trident. In this case we only have an intermediate photon, but no intermediate fermions, which means that to obtain the probability summed and averaged over the spins of initial and final state particles we only need to consider four terms, namely~\eqref{CaveR}, \eqref{CgammaR}, \eqref{BWaveR} and~\eqref{BWgammaR}.
It turns out to be convenient to write all spin and polarization sums as averages, so as our initial ansatz we take
\be\label{tridentGlue}
\mathbb{P}_{\rm glue}=\frac{2^4}{2}
\langle\mathbb{P}_{\rm C}\mathbb{P}_{\rm BW}\rangle+(1\leftrightarrow2) \;,
\ee  
where we have a factor of $2^3$ because we have replaced sums with averages over the spins for the three particles in the final state, we have similarly included a factor of $2$ for the intermediate photon, the factor of $1/2$ is because we have two identical particles in the final state and the $(1\leftrightarrow2)$ term makes the probability symmetric with respect to these two particles. Here and in the following we have, to avoid clutter, omitted the arguments of $\mathbb{P}_{\rm C}$ and $\mathbb{P}_{\rm BW}$ in $\langle\dots\rangle$, which are of course different. Apart from the different momenta and ${\bf n}$ we have chosen $\phi_1$ and $\phi_2$ for the first factor/step in $\langle\dots\rangle$, $\phi_3$ and $\phi_4$ for the second factor etc. In this gluing approach we also include step functions, e.g. $\theta(\sigma_{43}-\sigma_{21})$, so the different steps happen in the right chronological order, but this is done in exactly the same way as in~\cite{Dinu:2017uoj} so we leave it implicit in the following.     
The average/sum over the spins (and polarization in the general case) of initial and final state particles simply corresponds to sums over two antiparallel vectors, e.g. ${\bf n}_1=\pm{\bf n}_1^r$ where ${\bf n}_1^r$ is an arbitrary vector. So, we have $\langle1\rangle=1$ and $\langle{\bf n}\rangle=0$. However, things are nontrivial for the spins of intermediate particles. We can still use $\langle1\rangle=1$ and $\langle{\bf n}\rangle=0$ which give
\be\label{PtridentGlueComp1}
\mathbb{P}_{\rm glue}=2^3[\langle\mathbb{P}^{\rm C}\rangle\langle\mathbb{P}^{\rm BW}\rangle+{\bf P}_\gamma^{\rm C}\!\cdot\!\langle{\bf n}{\bf n}\rangle\!\cdot\!{\bf P}_\gamma^{\rm BW}+(1\leftrightarrow2)] \;,
\ee 
where ${\bf n}$ is the `polarization vector' of the intermediate photon. The question now is what to do with $\langle{\bf n}{\bf n}\rangle$. If one sums over ${\bf n}=\pm{\bf n}^{(r)}$, where ${\bf n}^{(r)}$ is some reference/basis vector, then the matrix $\langle{\bf n}{\bf n}\rangle$ clearly depends on the choice of ${\bf n}^{(r)}$.            
 We want $\mathbb{P}_{\rm glue}$ to be equal to $\mathbb{P}_2$ in~\eqref{P2start}, which we can write as
\be\label{PtridentComp2}
\mathbb{P}_2=2^3[\langle\mathbb{P}^{\rm C}\rangle\langle\mathbb{P}^{\rm BW}\rangle+{\bf P}_\gamma^{\rm C}\!\cdot\!{\bf P}_\gamma^{\rm BW}+(1\leftrightarrow2)] \;.
\ee
For linear polarization ${\bf a}_2(\phi)=0$ we have ${\bf P}_{\gamma,1}^{\rm C,BW}={\bf P}_{\gamma,2}^{\rm C,BW}=0$ and then we can obtain $\mathbb{P}_2$ by summing over two real polarization vectors with $\lambda=0$ and $\rho=0,\pi$, which correspond to ${\bf n}=\pm{\bf e}_3$. So, for linear polarization there is a choice of ${\bf n}^{(r)}$ that gives the desired result.
However, this does not work for more general field polarization where ${\bf P}_{\gamma,1}^{\rm C,BW},{\bf P}_{\gamma,2}^{\rm C,BW}\ne0$, because then we would not be able to obtain e.g. the term with $\frac{\kappa_{01}}{2}W_{12}\frac{\kappa_{23}}{2}W_{43}$ in $\mathbb{P}_2$. If one instead tried to sum over circular polarizations, with $\rho=\pi/2$ and $\lambda=\pm\pi/2$, then one would have ${\bf n}_3=0$ and would again be missing terms. The fact that we in general do not recover all terms in this way is because we already have a sum over the photon polarization on the amplitude level, so on the probability level we have in general a double sum, over $\epsilon$ and $\epsilon'$ say, and the naive gluing approach only takes into account the two terms where $\epsilon=\epsilon'$.

However, by comparing~\eqref{PtridentGlueComp1} and~\eqref{PtridentComp2} we immediately see that
by simply replacing $\langle{\bf n}{\bf n}\rangle\to{\bf 1}$ we obtain all terms in $\mathbb{P}_2$ for any polarization. We thus propose the following improved gluing approach:
Include a factor of $2$ for each intermediate particle and then simplify with the rules $\langle1\rangle=1$, $\langle{\bf n}\rangle=0$ and $\langle{\bf n}{\bf n}\rangle\to{\bf 1}$. We have showed in~\cite{Dinu:2018efz} that this procedure also works for double nonlinear Compton scattering, where we have a vector ${\bf n}$ for an intermediate electron rather than a photon. We have in fact showed~\cite{Dinu:2018efz} that this procedure also works for triple and quadruple nonlinear Compton scattering, where an electron interacts nonlinearly with the background field and emits three and four photons, respectively. These processes only have intermediate electrons and are hence built from~\eqref{CaveR}, \eqref{C0R}, \eqref{C1R} and~\eqref{C01R}.       

\begin{figure}
\centering
\includegraphics[width=0.95\linewidth]{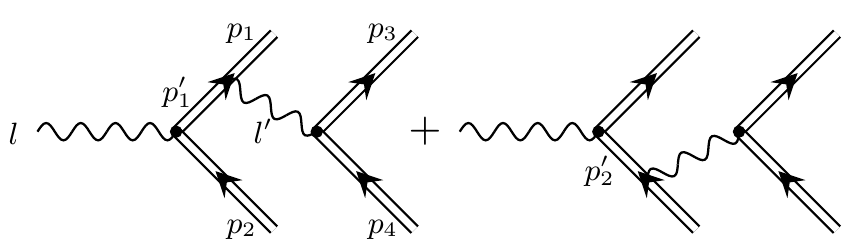}
\caption{This diagrams illustrates double nonlinear Breit-Wheeler pair production.}
\label{doubleBWdiagram}
\end{figure}

As an example of a process that involves both an intermediate photon and an intermediate fermion we consider double Breit-Wheeler pair production, i.e the decay of an initial photon into two electron-positron pairs as illustrated in Fig.~\ref{doubleBWdiagram}.
\begin{figure}
	\centering
	\includegraphics[width=0.95\linewidth]{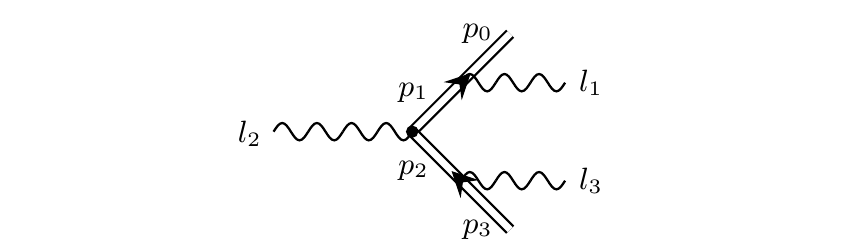}
	\caption{This diagrams illustrates one part of the process where one photon is emitted by the electron and another photon is emitted by the positron. This is an example with two cascade branches.}
	\label{twoBranchC}
\end{figure}
There are two different contributions to this process, where the intermediate photon is emitted by either an electron or a positron. We have checked that both can be obtained from
\be\label{doubleBWglue}
\begin{split}
\mathbb{P}_{\rm glue}^{\rm2BW}=&\frac{2^6}{2^2}[\langle\mathbb{P}_{\rm BW}\mathbb{P}_{\rm C}\mathbb{P}_{\rm BW}\rangle+\langle\mathbb{P}_{\rm BW}\mathbb{P}_{\rm C}^{\rm p}\mathbb{P}_{\rm BW}\rangle] \\
&+\text{permutations} \;,
\end{split}
\ee
where $\mathbb{P}_{\rm C}^{\rm p}$ is the probability of positron Compton scattering, we have $2^6$ because we have two intermediate particles and we sum over the spins of the final state particles and average over the initial polarization, and we have a factor of $1/2^2$ because there are two pairs of identical particles in the final state. The brackets in~\eqref{doubleBWglue} are calculated using $\langle1\rangle=1$, $\langle{\bf n}\rangle=0$ and $\langle{\bf n}{\bf n}\rangle={\bf 1}$ for all 7 ``spin/polarization vectors''.  
 
In Fig.~\ref{twoBranchC} we have (one part of) a process with two cascade branches. (The two photons can of course also be emitted by the same fermion, but that is another example of a single-branch cascade.) We have checked that the absolute squared of the diagram in Fig.~\ref{twoBranchC} can be obtained from
\be
\frac{2^6}{2}\langle\mathbb{P}_{\rm BW}\mathbb{P}_{\rm C}\mathbb{P}_{\rm C}^{\rm p}\rangle \;.
\ee  
So, our gluing approach is not restricted to single-branch cascades.

\subsection{Matrix approach}

Consider higher-order nonlinear Compton scattering (emission of more than one photon with nonlinear interaction with the background field). For this single-branch cascade we can replace the $\langle...\rangle$ ``operator'' with a matrix formulation. From the 3D spin unit vector ${\bf n}$ we define a 4D vector
\be
{\bf N}=\{1,{\bf n}\} \;.
\ee  
Summing over the photon polarization we find that the probability of single Compton scattering can be expressed as
\be
\mathbb{P}_{\rm C}=\langle\mathbb{P}\rangle+{\bf n}_0\!\cdot\!{\bf P}_0+{\bf P}_1\!\cdot\!{\bf n}_1+{\bf n}_0\!\cdot\!{\bf P}_{01}\!\cdot\!{\bf n}_1={\bf N}_1\!\cdot\!{\bf M}_{10}\!\cdot\!{\bf N}_0 \;,
\ee
with ${\bf N}_0$ and ${\bf N}_1$ for the initial and final state electron, respectively, and where ${\bf P}$ is defined in~\cite{Dinu:2018efz}.
The last step defines a $4\times4$ matrix ${\bf M}$ in terms of ${\bf P}_0$, ${\bf P}_1$ and ${\bf P}_{01}$. As an illustration of this matrix formulation, consider triple nonlinear Compton scattering. The $\langle...\rangle$ approach described in~\cite{Dinu:2018efz} can be expressed in terms of the 4D approach according to
\be
\begin{split}
\Big\langle
&[\langle\mathbb{P}\rangle+{\bf n}_0\!\cdot\!{\bf P}_0+{\bf P}_1\!\cdot\!{\bf n}_1+{\bf n}_0\!\cdot\!{\bf P}_{01}\!\cdot\!{\bf n}_1] \\
&[\langle\mathbb{P}\rangle+{\bf n}_1\!\cdot\!{\bf P}_0+{\bf P}_1\!\cdot\!{\bf n}_2+{\bf n}_1\!\cdot\!{\bf P}_{01}\!\cdot\!{\bf n}_2] \\
&[\langle\mathbb{P}\rangle+{\bf n}_2\!\cdot\!{\bf P}_0+{\bf P}_1\!\cdot\!{\bf n}_3+{\bf n}_2\!\cdot\!{\bf P}_{01}\!\cdot\!{\bf n}_3]\Big\rangle \\
=&\{1,{\bf 0}\}\!\cdot\!{\bf M}_{10}(3)\!\cdot\!{\bf M}_{10}(2)\!\cdot\!{\bf M}_{10}(1)\!\cdot\!\{1,{\bf 0}\} \;,
\end{split}
\ee
where ${\bf N}_0={\bf N}_1=\{1,{\bf 0}\}$ means that we are averaging and summing over the spin of the initial and final state electron, and ${\bf M}_{10}(i)$ depends on the lightfront time integration variables and longitudinal momenta associated with the emission of photon $i$. This seems like a simple formulation for a single-branch cascade, and it is similar to the M\"uller-matrix formulation of the evolution of Stokes vectors in optics, see also~\cite{McMasterRevModPhys} for perturbative QED with no background field.
However, in cascades with more than one branch, a $4\times4$ matrix would not describe the most general spin and polarization dependence, instead the dimensionality increases with the number of branches.   

We will now demonstrate that these two formulations are equivalent. The spin and polarization dependent parts factorize because for a fermion propagator we can express
\be
\slashed{p}+1=\sum_{\rho=\rho_0,\rho_0+\pi}u\bar{u}
\ee 
and
\be
\slashed{p}-1=\sum_{\rho=\rho_0,\rho_0+\pi}v\bar{v} \;,
\ee
where the spinors $u$ and $v$ are given by~\eqref{generalu} and~\eqref{generalv}. And for the photon propagator we have
\be
L_{\mu\nu}=g_{\mu\nu}-\frac{1}{kl}(k_\mu l_\nu+l_\mu k_\nu)=\sum_{\rho=\rho_0,\rho_0+\pi}\epsilon_\mu\bar{\epsilon}_\nu \;,
\ee
where the polarization vector $\epsilon_\mu$ is given by~\eqref{epsfromrholambda}. Consider a Compton scattering step, where a photon is emitted by an electron. From the amplitude we have, regardless of whether or not the particles come from the initial, some intermediate or the final state,
\be
\mathcal{X}:=\bar{\psi}(p_n,\rho_n,\lambda_n,\phi_n)\bar{\slashed{\epsilon}}(l,\rho_l,\lambda_l)\psi(p_m,\rho_m,\lambda_m,\phi_n) \;,
\ee
where
\be
\psi(p,\rho,\lambda,\phi)=K(p,\phi)u(p,\rho,\lambda)\varphi(p,\phi) \;,
\ee
\be
K=1+\frac{\slashed{k}\slashed{a}}{2kp}
\ee
and
\be
\varphi=\exp\left\{-i\left(px+\int^\phi\frac{2ap-a^2}{2kp}\right)\right\} \;.
\ee
From the complex conjugate of the amplitude we have
\be
\mathcal{Y}:=\bar{\psi}(p_m,\rho_m^c,\lambda_m^c,\phi_m)\slashed{\epsilon}(l,\rho_l^c,\lambda_l^c)\psi(p_n,\rho_n^c,\lambda_n^c,\phi_m) \;,
\ee 
where $\rho_i^c$ and $\lambda_i^c$ can be different from $\rho_i$ and $\lambda_i$ for intermediate particles. We will integrate $\mathcal{X}\mathcal{Y}$ over the transverse momenta. Assume first that we are dealing with a final-state step, where no more particles are produced by these particles (but more particles can be produced at a later lightfront time by a different cascade branch). We have three cases: 1) We use the overall momentum conservation delta function to perform the $p_n^\LCperp$ integral, which means $p_n^\LCperp=p_m^\LCperp-l^\LCperp$, where $p_m^\LCperp$ depends on the other, independent momenta. Only this step depends on $l^\LCperp$, so we have
\be\label{intXY}
\mathcal{Z}=\mathcal{N}\int\ud^2l^\LCperp \mathcal{X}\mathcal{Y} \;,
\ee   
where $\mathcal{N}$ is a normalization factor, which we will come back to.
2) If we have already used the overall delta function for a different step, then we have integrals over both $l^\LCperp$ and $p_n^\LCperp$. These momenta also appear in some other step(s) because of the momentum conservation, but only via their sum $p_m^\LCperp=p_n^\LCperp+l^\LCperp$. So, we change variable from $p_n^\LCperp$ to $p_m^\LCperp$, and then $l^\LCperp$ only appears in this step and we again have~\eqref{intXY}. 3) If we initially have an integral over $p_n^\LCperp$ and where the photon momentum is fixed by momentum conservation $l^\LCperp=p_m^\LCperp-p_n^\LCperp$, then we just change variable from $p_n^\LCperp$ to $l^\LCperp$.
So, in all cases we have~\eqref{intXY}. After performing the integral over $l^\LCperp$ we find that the result is independent of all the other transverse momenta. This means that this step factors out and we can treat the step that produced the $p_m$ electron as if it were a final step. After performing all the transverse momentum integrals we find that all steps have factorized. The different steps are linked via matrix multiplication. 
For an initial electron we have $u_\alpha(\rho,\lambda)\bar{u}_\beta(\rho,\lambda)$, with $u$ from the amplitude and $\bar{u}$ from its complex conjugate, which we can write in terms of the basis spinors as (omitting the spinor indices)
\be
u\bar{u}=N^{(0)}\cdot(u_{\scriptscriptstyle\uparrow}\bar{u}_{\scriptscriptstyle\uparrow},u_{\scriptscriptstyle\uparrow}\bar{u}_{\scriptscriptstyle\downarrow},u_{\scriptscriptstyle\downarrow}\bar{u}_{\scriptscriptstyle\uparrow},u_{\scriptscriptstyle\downarrow}\bar{u}_{\scriptscriptstyle\downarrow}) \;,
\ee
where
\be
N^{(0)}=\Big(\cos^2\frac{\rho}{2},\cos\frac{\rho}{2}\sin\frac{\rho}{2} e^{-i\lambda},\cos\frac{\rho}{2}\sin\frac{\rho}{2} e^{i\lambda},\sin^2\frac{\rho}{2}\Big) \;.
\ee
For an outgoing electron we have $\bar{u}_\alpha(\rho,\lambda)u_\beta(\rho,\lambda)$, with $\bar{u}$ from the amplitude and $u$ from its complex conjugate, so
\be
\bar{u}u=\bar{N}^{(0)}\cdot(\bar{u}_{\scriptscriptstyle\uparrow}u_{\scriptscriptstyle\uparrow},\bar{u}_{\scriptscriptstyle\uparrow}u_{\scriptscriptstyle\downarrow},\bar{u}_{\scriptscriptstyle\downarrow}u_{\scriptscriptstyle\uparrow},\bar{u}_{\scriptscriptstyle\downarrow}u_{\scriptscriptstyle\downarrow}) \;.
\ee
Similarly, for an outgoing photon we have $\bar{\epsilon}_\mu\epsilon_\nu$, with the first term from the amplitude and the second term from its complex conjugate,
\be
\bar{\epsilon}\epsilon=\bar{N}^{(0)}\cdot(\bar{\epsilon}_\uparrow\epsilon_\uparrow,\bar{\epsilon}_\uparrow\epsilon_\downarrow,\bar{\epsilon}_\downarrow\epsilon_\uparrow,\bar{\epsilon}_\downarrow\epsilon_\downarrow) \;,
\ee
where $\epsilon_\uparrow^\mu$ is given by $(\rho=0,\lambda=0)$ and $\epsilon_\downarrow^\mu$ by $(\rho=\pi,\lambda=0)$.
For an intermediate particle we a double spin sum, one for the amplitude and a second for its complex conjugate. We can without loss of generality sum over the spin basis with $\lambda=0$, i.e. $\sum_{r,s=\uparrow,\downarrow}u_r\bar{u}_s$. We can make this a single sum by summing over $(u_{\scriptscriptstyle\uparrow}\bar{u}_{\scriptscriptstyle\uparrow},u_{\scriptscriptstyle\uparrow}\bar{u}_{\scriptscriptstyle\downarrow},u_{\scriptscriptstyle\downarrow}\bar{u}_{\scriptscriptstyle\uparrow},u_{\scriptscriptstyle\downarrow}\bar{u}_{\scriptscriptstyle\downarrow})$. We order the sums for the other two particles also as $(\uparrow\uparrow,\uparrow\downarrow,\downarrow\uparrow,\downarrow\downarrow)$. The most general case is thus given by a $4\times4\times4$ matrix $\mathcal{Z}_{i_li_mi_n}$, where $i_j=1,..,4$ and the $\{$first, second, third$\}$ index for the $\{$photon, incoming electron, outgoing electron$\}$. We want to express the spin and polarization of the particles in terms of their Stokes vectors (as above) rather than $N^{(0)}$, and for this it is natural to transform the indices using
\be
\mathcal{T}:=\frac{1}{2}\begin{pmatrix}1&0&0&1\\0&1&-i&0\\0&1&i&0\\1&0&0&1\end{pmatrix} \;,
\ee
which obeys $2\mathcal{T}\mathcal{T}^\dagger=1$ ($\mathcal{T}^\dagger$ is the conjugate transpose). Using this matrix we define
\be
\mathcal{M}_{i_li_mi_n}=\mathcal{T}^\dagger_{i_lj_l}\mathcal{T}^\dagger_{i_nj_n}\mathcal{Z}_{j_lj_mj_n}\mathcal{T}_{j_mi_m} \;.
\ee
If one of these particles is in the initial or final state, then we have
\be
2\mathcal{T}^\dagger N^{(0)}=(1,\cos\lambda\sin\rho,\sin\lambda\sin\rho,\cos\rho)=N 
\ee
and $\bar{N}^{(0)}2\mathcal{T}=N$,
which is the Stokes vector as defined above. For single Compton scattering we have
\be
N^l_{i_l}N^n_{i_n}\mathcal{M}_{i_li_mi_n}N^m_{i_m} \;,
\ee
where $N^l$, $N^m$ and $N^n$ are the Stokes vectors for the photon and the initial and final state electron, respectively. So, the building block $\mathcal{M}$ for constructing the gluing estimate of higher orders is naturally interpreted in terms of the Stokes-vector dependent first-order process. For double Compton scattering, for example, we would have
\be
N^{l_2}_{i_{l_2}}N^{l_1}_{i_{l_1}}N^n_{i_n}\mathcal{M}_{i_{l_2}i_{m_1}i_n}\mathcal{M}_{i_{l_1}i_{m_0}i_{m_1}}N^m_{i_{m_0}} \;.
\ee  

We can find similar expressions for photon emission by a positron and pair production. Because an incoming positron has $\bar{v}$ instead of $v$ in the amplitude, for the positron spin sums we order the terms according to $(\uparrow\uparrow,\downarrow\uparrow,\uparrow\downarrow,\downarrow\downarrow)$ instead of $(\uparrow\uparrow,\uparrow\downarrow,\downarrow\uparrow,\downarrow\downarrow)$, so that we have the same $N^{(0)}$ and $\mathcal{T}$ for all particles.

To show that this matrix formulation is equivalent to the average $\langle...\rangle$ approach, start with the gluing ansatz 
\be
\langle...M\!\cdot\!NN\!\cdot\!M...\rangle \;,
\ee
where $N$ is the Stokes vector of some intermediate particle. Using the prescription $\langle1\rangle=1$, $\langle{\bf n}\rangle=0$ and $\langle{\bf n}{\bf n}\rangle={\bf 1}$ we also have $\langle N_\alpha N_\beta\rangle=\delta_{\alpha\beta}$, so the $\langle...\rangle$ approach reduces to the matrix formulation.

We now return to the normalization factor $\mathcal{N}$ in~\eqref{intXY}. We use the same normalization as in~\cite{Dinu:2017uoj}. So, in particular, the amplitude is given by
\be
\frac{1}{k_\LCp}\bar{\delta}(P_{\rm in}-P_{\rm out})M:=\bra{0}\prod_{e^\LCm}b\prod_{e^\LCp}d\prod_{\gamma}a\; U b^\dagger\ket{0} \;,
\ee
where $b$, $d$ and $a$ are the mode operators (with the momentum and spin arguments suppressed) for electrons, positrons and photons, respectively, and $U$ is the evolution operator. The delta function is given by $\bar{\delta}(P)=(2\pi)^3\delta(P_\LCm)\delta^2(P_\LCperp)$.
$P_{\rm in}$ is the momentum of the initial particle and $P_{\rm out}$ is the sum of the ($\LCm$ and $\LCperp$ components of the) momenta of all the final state particles. The initial state is given by
\be
\ket{\rm in}=\int\ud\tilde{P}_{\rm in}f(P_{\rm in})B^\dagger\ket{0} \;,
\ee    
where $B$ is the mode operator for an electron, a positron or a photon, and $\ud\tilde{P}_{\rm in}=\theta(P^{\rm in}_\LCm)\ud P^{\rm in}_\LCm\ud^2 P^{\rm in}_\LCperp/(2P^{\rm in}_\LCm(2\pi)^3)$ (which is Lorentz invariant). The mode operators are normalized such that $\bracket{\rm in}{\rm in}=1$ implies
\be
\int\ud\tilde{P}_{\rm in}|f|^2=1 \;.
\ee
We assume for simplicity that the wave packet is sharply peaked, which means
\be\label{Pinpp}
\int\ud\tilde{p}'\left|\int\ud\tilde{P}_{\rm in}f\frac{1}{k_\LCp}M\right|^2=\frac{\theta(kp')}{kP_{\rm in}kp'}|M|^2 \;,
\ee
where $p'$ is the momentum of one of the final-state particles. For each of the rest of the outgoing particles we have a momentum integral $\int\ud\tilde{P}$. For each intermediate particle we have $i\int\ud^4x\ud^4P=i\int\ud\phi\frac{\ud P_\LCp}{2\pi k_\LCp}\ud^3x^{\LCm,\LCperp}\ud^3P_{\LCm,\LCperp}$. The integral over $x^{\LCm,\LCperp}$ gives a delta function which we use to perform the integral over $P_{\LCm,\LCperp}$. The $P_\LCp$ integral is elementary and independent of the field, 
\be
\frac{i}{k_\LCp}\int\frac{\ud P_\LCp}{2\pi}\frac{(1,P_\LCp)}{P^2-m^2+i\epsilon}e^{-iP\Delta x^\LCp} \;,
\ee    
where $\Delta x^\LCp$ is the difference between two $x^\LCp$ variables, and $(1,P_\LCp)$ means that the numerator is either independent of or linear in $P_\LCp$. Using $P^2=4P_\LCm P_\LCp-P_\LCperp^2$, the integral either gives a term with an ``instantaneous'' $\delta(\Delta x^\LCp)$ or one with a time-ordering $\theta(\Delta x^\LCp)$, cf.~\cite{Seipt:2012tn}. Only the step-function term contributes to the cascade/gluing estimate. Apart from the terms that we have already included in $\mathcal{X}$ and $\mathcal{Y}$, each propagator gives a factor of $1/(2kP)$. There is one $x^\mu$ integral more than there are propagators and, since $\ud^4x=\ud x^\LCp\ud x^\LCm\ud^2 x^\LCperp/2$, it gives an overall factor of $1/4$, which we combine with $1/(kP_{\rm in}kp')$ in~\eqref{Pinpp}.   
Thus, each external and intermediate particle gives a factor of $1/(2kP)$. So, in order for 
\be
\mathbb{P}=\int\prod\ud kP_i\theta(kp')\prod\ud\phi_i\text{``time-ordering''}\prod_\text{all steps}\mathcal{Z}_i 
\ee
to give the probability with the correct normalization,
where $kP_i=b_0s_i$ are all the independent longitudinal momentum variables of the outgoing particles and ``time-ordering'' denote the product of step functions that give lightfront-time ordering, we need
\be
\mathcal{N}=\frac{e^2}{(2\pi)^3}\frac{1}{2kp_m}\frac{1}{2kp_n}\frac{1}{2kl} \;.
\ee  

The fundamental matrix for a Compton-scattering step can be expressed compactly as 
\be
\mathcal{M}^{\rm C}=\frac{i\alpha}{8\pi b_0^2s_m^2\theta_{nm}}\hat{\mathcal{M}}^{\rm C} \;,
\ee
\be
\begin{split}
\hat{\mathcal{M}}^{\rm C}_{1i_mi_n}=&\frac{q^2\sigma^{03}_3}{2s_ms_n}+C_1-\frac{\tilde{\kappa}}{2}{\bf w}_n\!\cdot\!\sigma_2\!\cdot\!{\bf w}_m\sigma^{01}_1 \\
&+\left[\frac{2ib_0}{r\theta_{nm}}+{\bf w}_n\!\cdot\!{\bf w}_m\right]\left[\frac{\kappa}{2}\sigma^{03}_0+\sigma^{12}_0\right]
\\
\hat{\mathcal{M}}^{\rm C}_{2i_mi_n}=&-\frac{q^2\sigma^{12}_1}{2s_ms_n}+C_2-\frac{\tilde{\kappa}}{2}{\bf w}_n\!\cdot\!\sigma_3\!\cdot\!{\bf w}_m i\sigma^{12}_2 \\
&+{\bf w}_n\!\cdot\!\sigma_1\!\cdot\!{\bf w}_m\left[\sigma^{03}_0+\frac{\kappa}{2}\sigma^{12}_0\right]
\\
\hat{\mathcal{M}}^{\rm C}_{3i_mi_n}=&-\frac{q^2i\sigma^{03}_2}{2s_ms_n}+C_3+\frac{\tilde{\kappa}}{2}\left[\frac{2ib_0}{r\theta_{nm}}+{\bf w}_n\!\cdot\!{\bf w}_m\right]\sigma^{03}_1 \\
&-{\bf w}_n\!\cdot\!\sigma_2\!\cdot\!{\bf w}_m\left[\frac{\kappa}{2}\sigma^{03}_0+\sigma^{12}_0\right]
\\
\hat{\mathcal{M}}^{\rm C}_{4i_mi_n}=&-\frac{q^2\sigma^{12}_3}{2s_ms_n}+C_4+\frac{\tilde{\kappa}}{2}{\bf w}_n\!\cdot\!\sigma_1\!\cdot\!{\bf w}_m i\sigma^{12}_2 \\
&+{\bf w}_n\!\cdot\!\sigma_3\!\cdot\!{\bf w}_m\left[\sigma^{03}_0+\frac{\kappa}{2}\sigma^{12}_0\right]
\end{split}
\ee
where $r=(1/s_n)-(1/s_m)$, $\kappa=(s_m/s_n)+(s_n/s_m)$, $\tilde{\kappa}=(s_m/s_n)-(s_n/s_m)$,
\be
\sigma^{03}_0=\begin{pmatrix}1&0&0&0\\0&0&0&0\\0&0&0&0\\0&0&0&1\end{pmatrix} \quad 
\sigma^{03}_1=\begin{pmatrix}0&0&0&1\\0&0&0&0\\0&0&0&0\\1&0&0&0\end{pmatrix} \quad \text{etc.}\;,
\ee
\be
\sigma^{12}_0=\begin{pmatrix}0&0&0&0\\0&1&0&0\\0&0&1&0\\0&0&0&0\end{pmatrix} \quad 
\sigma^{12}_1=\begin{pmatrix}0&0&0&0\\0&0&1&0\\0&1&0&0\\0&0&0&0\end{pmatrix} \quad \text{etc.}\;,
\ee
\be
C_1=\begin{pmatrix}0&\frac{q}{s_n}V_1&\frac{q}{s_n}V_2&0\\ \frac{q}{s_m}V_1&0&0&-\frac{q}{s_m}X_1\\ \frac{q}{s_m}V_2&0&0&-\frac{q}{s_m}X_2\\0&\frac{q}{s_n}X_1&\frac{q}{s_n}X_2&0\end{pmatrix}  \;,
\ee
\be
C_2=-\begin{pmatrix}0&\frac{q}{s_m}V_2&\frac{q}{s_m}V_1&0\\ \frac{q}{s_n}V_2&0&0&\frac{q}{s_n}X_2\\ \frac{q}{s_n}V_1&0&0&\frac{q}{s_n}X_1\\0&-\frac{q}{s_m}X_2&-\frac{q}{s_m}X_1&0\end{pmatrix}  \;,
\ee
\be
C_3=\begin{pmatrix}0&\frac{q}{s_n}X_1&\frac{q}{s_n}X_2&0\\ \frac{q}{s_m}X_1&0&0&-\frac{q}{s_m}V_1\\ \frac{q}{s_m}X_2&0&0&-\frac{q}{s_m}V_2\\0&\frac{q}{s_n}V_1&\frac{q}{s_n}V_2&0\end{pmatrix}  \;,
\ee
\be
C_4=\begin{pmatrix}0&-\frac{q}{s_m}V_1&\frac{q}{s_m}V_2&0\\-\frac{q}{s_n}V_1&0&0&-\frac{q}{s_n}X_1\\ \frac{q}{s_n}V_2&0&0&\frac{q}{s_n}X_2\\0&\frac{q}{s_m}X_1&-\frac{q}{s_m}X_2&0\end{pmatrix}  \;,
\ee
with ${\bf X}$ and ${\bf V}$ obtained from~\eqref{XVdef} by replacing ${\bf w}_1\to{\bf w}_m$ and ${\bf w}_2\to{\bf w}_n$.
For single Compton scattering,
$N^l_{i_l}N^n_{i_n}\mathcal{M}^{\rm C}_{i_li_mi_n}N^m_{i_m}$
is equivalent to~\eqref{PCnnn}, \eqref{PCfromRC} to~\eqref{Cgamma01R}. The corresponding matrices for positron Compton and Breit-Wheeler, $\mathcal{M}^{\rm pC}$ and $\mathcal{M}^{\rm BW}$, can be obtained from $\mathcal{M}^{\rm C}$ using the same replacement rules as explained for~\eqref{PCfromRC} to~\eqref{Cgamma01R}.

\subsection{Spin dependence}\label{SpinSection}

The spin/polarization treatment above is needed in order to sum over the intermediate particles. The same treatment can also be used in order to study the dependence on the spin/polarization of initial and final state particles.
In this section we consider the dependence of the two-step part of trident on the spin of the initial electron, described by the vector ${\bf n}$ as in Sec.~\ref{gluing-section}. 
Spin effects in single and double nonlinear Compton scattering have been studied in e.g.~\cite{King:2014wfa,DelSorbo:2017fod,Seipt:2018adi}.
By either a direct calculation or by simply omitting the average over ${\bf n}$ in $\langle\mathbb{P}_{\rm C}\mathbb{P}_{\rm BW}\rangle$ 
we find 
\be
\mathbb{P}_{\rm two}=\langle\mathbb{P}\rangle+{\bm n}\!\cdot\!{\bm P} \;,
\ee
where the first term gives the spin average,
which we calculated in~\cite{Dinu:2017uoj},
and the second term gives the spin dependence, where ${\bm P}={\bm P}_\LCperp+{\bm P}_{\scriptscriptstyle\parallel}$,  
\be\label{PspinPerp}
\begin{split}
	{\bm P}_\LCperp=\frac{i\alpha^2}{8\pi^2b_0^2}&\int\!\ud^4\phi\frac{\theta(\theta_{31})\theta(\theta_{42})}{q_1\theta_{21}\theta_{43}}e^{\frac{i}{2b_0}\left[r_1\Theta_{21}+r_2\Theta_{43}\right]} \\ 
	&\bigg\{-\frac{\kappa_{23}}{2}W_{34}{\bf X}
	+\bigg({\bm\sigma}_1\frac{H_2}{s_1}-{\bm\sigma}_3\frac{C_2}{s_1}+ \\
	&i{\bm\sigma}_2\left[\frac{\kappa_{23}}{2}\left(\frac{2i b_0}{r_2\theta_{43}}+1+D_2\right)+1\right]\bigg)\!\cdot\!{\bm Y}\bigg\}  \\
	&+(s_1\leftrightarrow s_2)
\end{split}
\ee
and 
\be\label{PspinPara}
\begin{split}
	{\bm P}_{\scriptscriptstyle\parallel}=&\frac{i\alpha^2}{8\pi^2b_0^2}\int\!\ud^4\phi\frac{\theta(\theta_{31})\theta(\theta_{42})}{q_1\theta_{21}\theta_{43}}e^{\frac{i}{2b_0}\left[r_1\Theta_{21}+r_2\Theta_{43}\right]} \\
	&\bigg\{\frac{1}{2}\left(\frac{1}{s_1}+1\right)W_{12}\left[\frac{\kappa_{23}}{2}\left(\frac{2ib_0}{r_2\theta_{43}}+1+D_2\right)+1\right]
	- \\
	&\left[\frac{1}{2}\left(\frac{1}{s_1}+1\right)\left(\frac{2ib_0}{r_1\theta_{21}}+1+D_1\right)-1\right]\frac{\kappa_{23}}{2}W_{34}\bigg\}\hat{\bf k} \\
	&+(s_1\leftrightarrow s_2) \;,
\end{split}
\ee  
where ${\bm X}=\frac{1}{2}({\bf w}_2+{\bf w}_1)$, ${\bm Y}=\frac{1}{2}({\bf w}_2-{\bf w}_1)$,
\be
H_2={\bm w}_3\!\cdot\!{\bm\sigma}_3\!\cdot\!{\bm w}_4={\bf w}_{31}{\bf w}_{41}-{\bf w}_{32}{\bf w}_{42}
\ee
\be 
C_2={\bm w}_3\!\cdot\!{\bm\sigma}_1\!\cdot\!{\bm w}_4={\bf w}_{31}{\bf w}_{42}+{\bf w}_{32}{\bf w}_{41}
\ee
and $W_{34}={\bm w}_3\cdot i{\bm\sigma}_2\cdot{\bm w}_4$.
Note that only the photon emission part of the integrand depends on the spin.

Consider a constant field $a_\mu=\delta_\mu^1 a_0\phi$. To obtain the dominant contribution we replace $\theta(\theta_{42})\theta(\theta_{31})\to\theta(\sigma_{43}-\sigma_{21})$. We find
\be
\begin{split}
	{\bm P}(s)=&{\bm e}_2\frac{\alpha^2(a_0\Delta\phi)^2}{2\chi^2q_1}\frac{{\rm Ai}\left(\xi_1\right)}{\sqrt{\xi_1}} \\
	&\left\{\left[\kappa_{23}-\frac{1}{s_1}\right]\frac{{\rm Ai}'(\xi_2)}{\xi_2}-{\rm Ai}_1(\xi_2)\right\} \;,
\end{split}
\ee
where $\xi_i=\left[r_i/\chi\right]^\frac{2}{3}$.
So the maximum and minimum probability is obtained with spin orthogonal to the field and the propagation direction.
For $\chi\ll1$ we find
\be
{\bm P}=-\frac{\chi}{27}{\bm e}_2\frac{\alpha^2(a_0\Delta\phi)^2}{64}\exp\left\{-\frac{16}{3\chi}\right\} \;,
\ee
so the spin dependence is smaller than the average by a factor of $\chi/27\ll1$ in this regime.
We recognize this factor from Eq.~(24) in~\cite{Ritus:1972nf}.

\section{Trident in a circularly polarized field}

In~\cite{Dinu:2019wdw} we demonstrated that our gluing method indeed gives a good approximation for trident in a long pulse with $a_0\sim1$ and linear polarization. Since the expressions presented above are valid for any field polarization, we will demonstrate this fact here for trident in a circularly polarized field,
\be\label{circularPolDefinition}
{\bm a}(\phi)=\frac{a_0}{\sqrt{2}}\{\sin(\phi),\cos(\phi),0\}e^{-(\phi/\mathcal{T})^2} \;.
\ee
For linear polarization the two-step part can be obtained by summing over a certain polarization basis for the intermediate photon, which means that the polarization aspect of the gluing approach is similar to the standard LCF case. For other field polarizations it is in general not possible to obtain the two-step by such a simple sum over two constant polarization vectors. So, for the field considered in this section our gluing approach is indispensable.   

In the following we compare the two-step part with the one-step terms, i.e. the rest of the probability. All the relevant definitions are given in~\cite{Dinu:2019wdw}. As shown in~\cite{Dinu:2019wdw}, for a very short pulse ($\mathcal{T}\sim1$), low to moderate intensity ($a_0\lesssim1$), and ``low'' energy ($b_0\lesssim1$), the exchange part of the one-step (i.e. the cross term between the two terms on the amplitude level that are related by exchanging the two identical particles in the final state) can be comparable to the other terms. This is in fact what one should expect if all parameters are on the order of one. This provides a numerical challenge as the exchange term is numerically intensive to compute. Fortunately, the exchange term becomes comparably small for long pulses ($\mathcal{T}\gg1$) and/or high intensity ($a_0\gg1$). 
So, since we focus on a relatively long pulse with $\mathcal{T}=80$, we have approximated the exchange term by its LCF version. We expect this to be good enough for $a_0=1$ and very good for higher $a_0$.  

\begin{figure*}
\makebox[\textwidth][c]{
\includegraphics[width=1.1\linewidth]{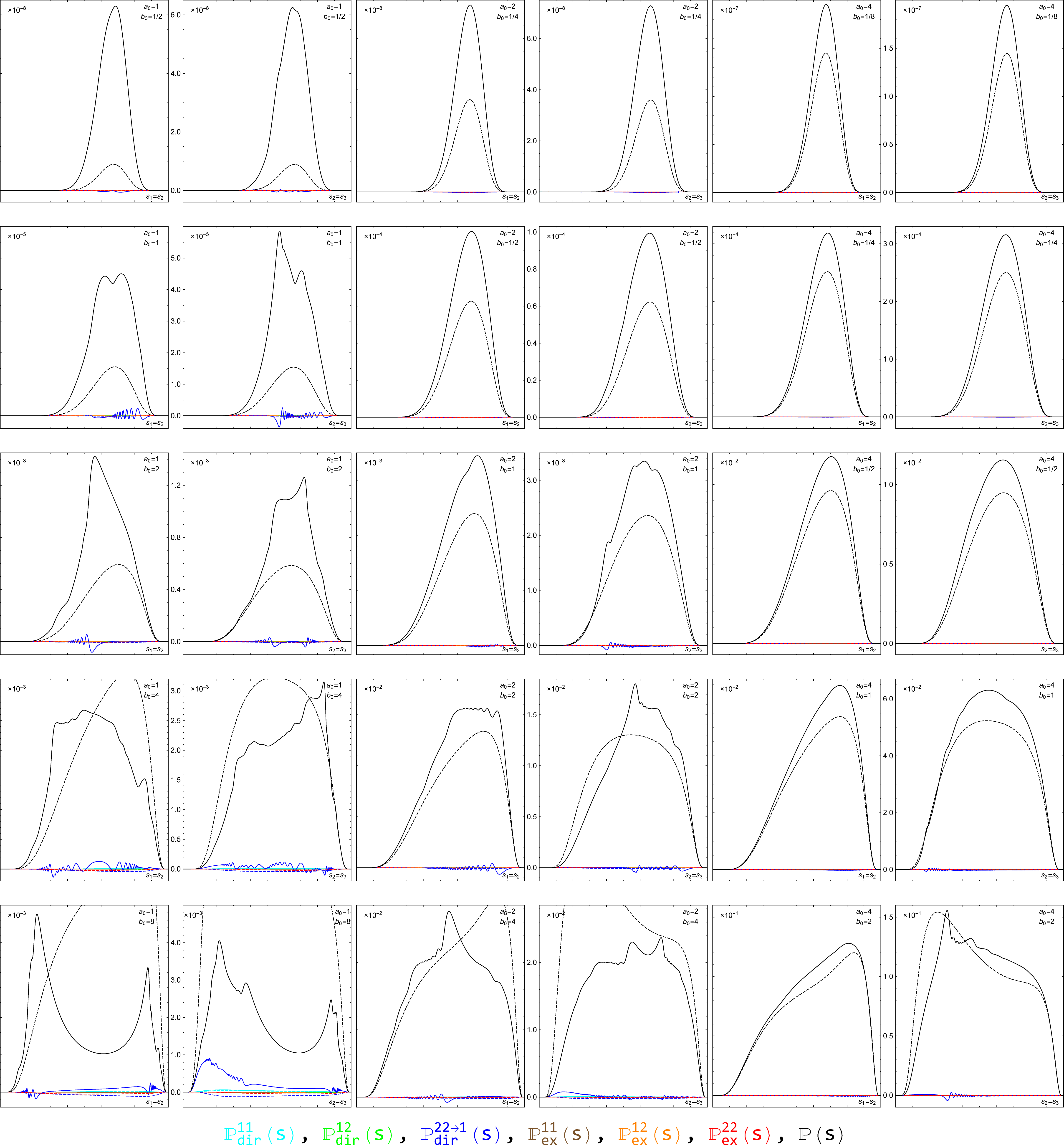}
}
\caption{Sections of the spectrum for $a_0=1,2,4$, $\chi=1/2,1,2,4,8$ and $\mathcal{T}=80$. Solid lines are exact and dashed lines show LCF. }
\label{sectionsFig1}
\end{figure*}

\begin{figure*}
\makebox[\textwidth][c]{
\includegraphics[width=1.1\linewidth]{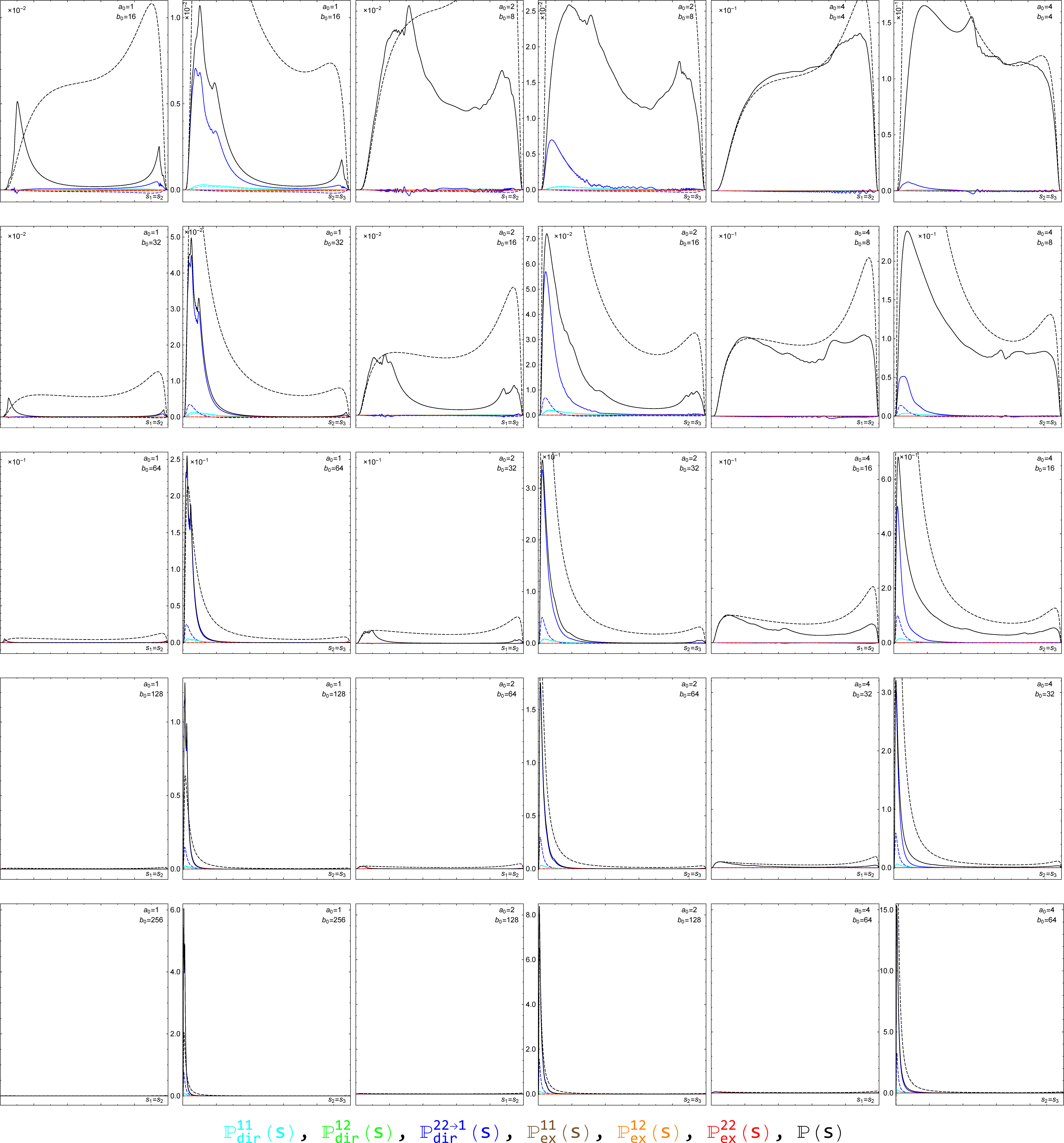}
}
\caption{Sections of the spectrum for $a_0=1,2,4$, $\chi=16,32,64,128,256$ and $\mathcal{T}=80$.}
\label{sectionsFig2}
\end{figure*}

\begin{figure*}
\vspace{-.5cm}
\includegraphics[width=.95\linewidth]{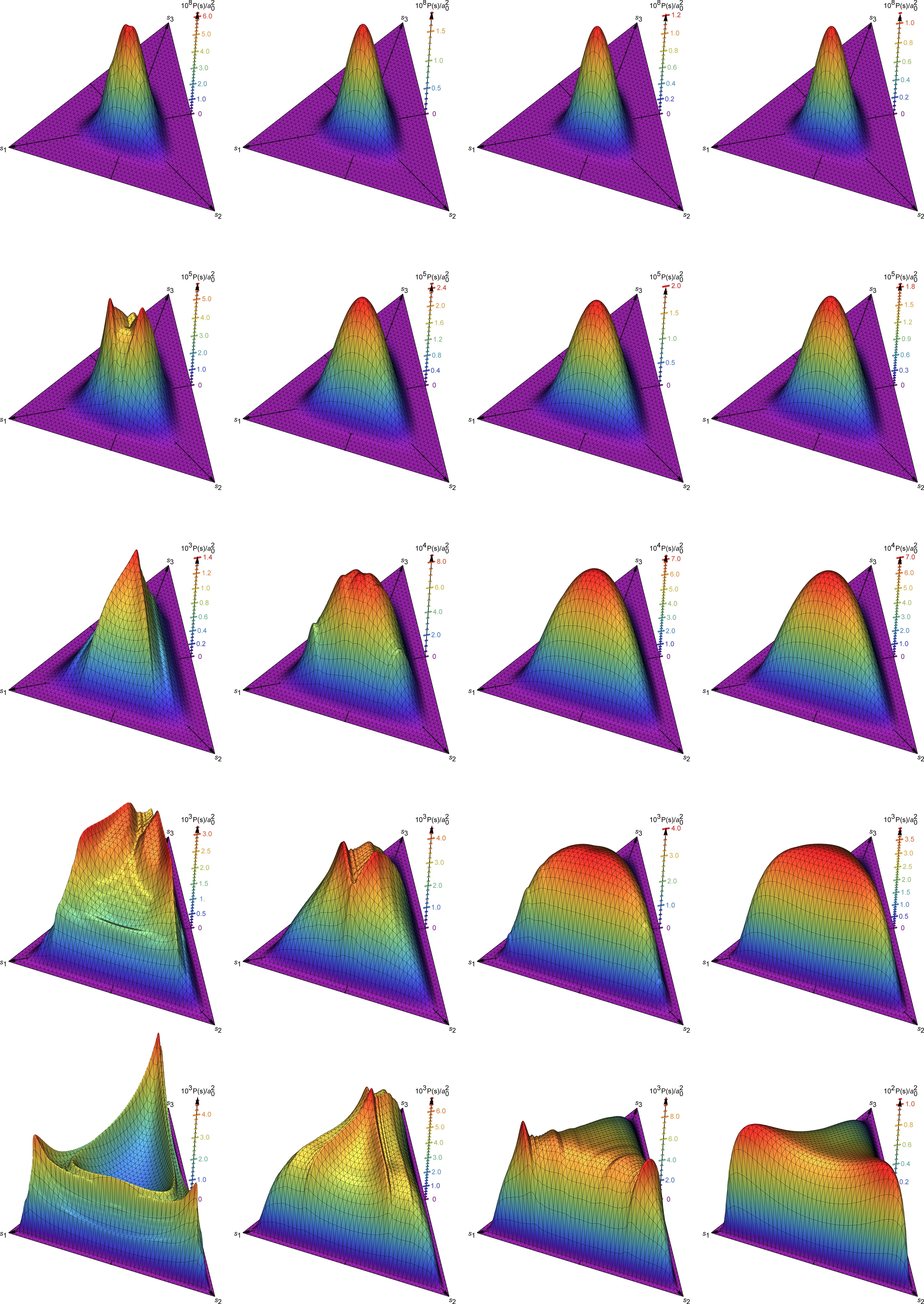}
\caption{The spectrum for $\mathcal{T}=80$, $a_0=1,2,4,8$ from left to right, and $\chi=1/2,1,2,4,8$ from top to bottom.}
\label{3DfigLowb0}
\end{figure*} 

\begin{figure*}
\vspace{-.5cm}
\includegraphics[width=.95\linewidth]{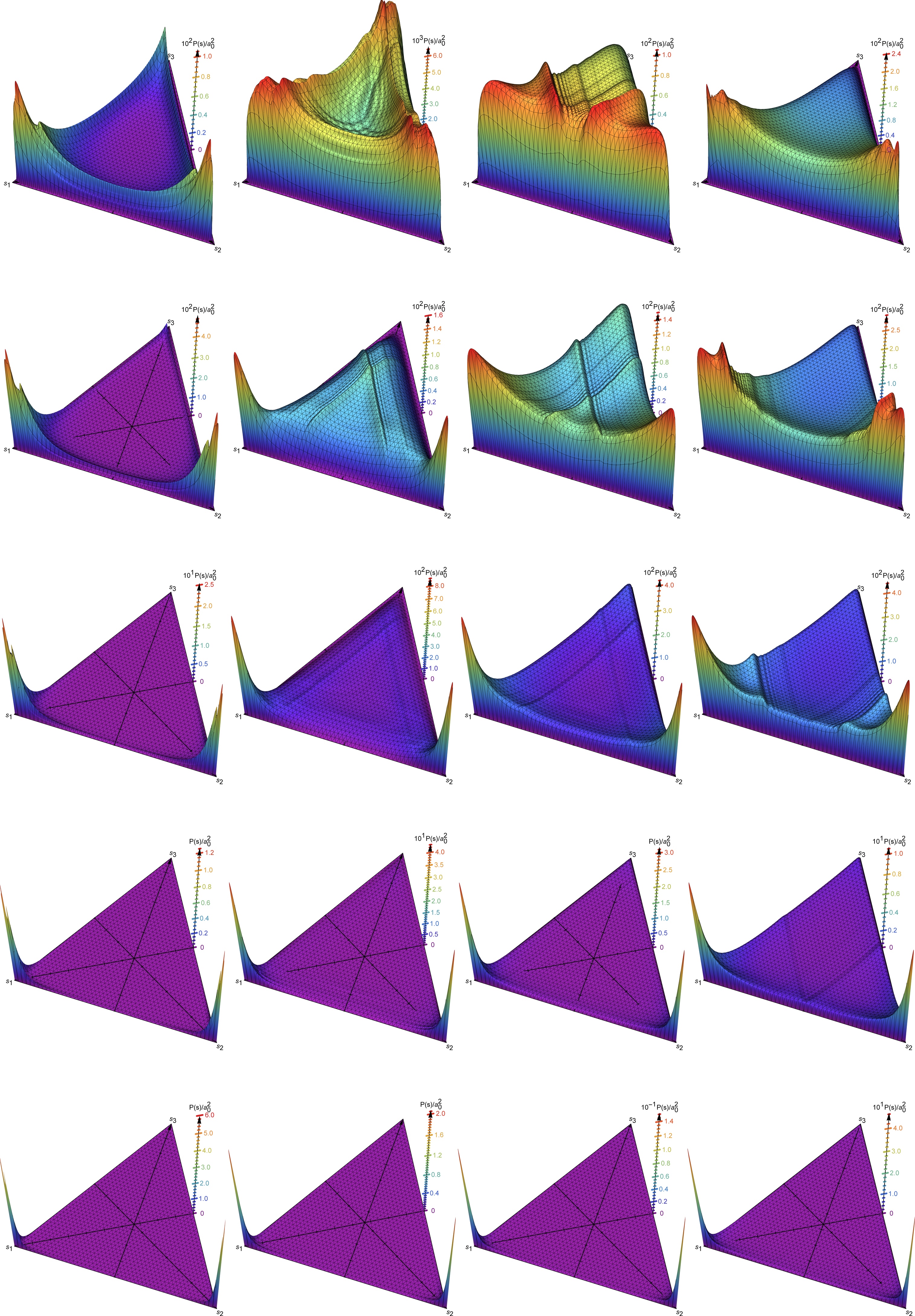}
\caption{Same as Fig.~\ref{3DfigLowb0} but with $\chi=16,32,64,128,256$ from top to bottom.}
\label{3DfigHighb0}
\end{figure*}

In Fig.~\ref{sectionsFig1} and Fig.~\ref{sectionsFig2} we show the two sections $s_1=s_2$ and $s_2=s_3$ for different values of $a_0$ and $\chi$, and for a long pulse with $\mathcal{T}=80$.
These plots contain curves for the 6 different contributions to the one-step part, as defined in~\cite{Dinu:2017uoj,Dinu:2019wdw}. However, here it is enough to note that $\mathbb{P}_{\rm ex}$ is the exchange term, which is negligible, and $\mathbb{P}_{\rm dir}^{22\to1}$ is a part of the one-step that can be obtained with our gluing method.
These plots show that our gluing method (which gives the two-step part of this process) gives indeed a good approximation. The one-step terms have some oscillations in the momentum spectrum, making the two-step approximation better for the integrated probability.  
Note that, while the LCF version breaks down when $a_0$ is not large, here we see that our generalized two-step is good also at moderate intensities, $a_0\sim1$. 
Comparing the exact result and the LCF approximation, we note that at low $b_0$ the LCF approximation is significantly smaller, then as $b_0$ increases it temporarily becomes much larger, and finally for large $b_0$ it again becomes smaller. 
This is because at large $b_0$, the LCF approximation is much larger than the exact result for the two-step term, but smaller than the exact value for the one-step terms. At moderately large $b_0$ this makes the total probability larger in the LCF approximation, but above a certain $b_0$ value, due to the dominance of the one-step, the exact total probability surpasses its LCF approximation. Of course, this happens sooner for lower $a_0$, where the two-step loses importance faster.
So, our gluing approach works in a significantly larger region of parameter space compared to LCF. These results also confirm the fact that our gluing method works for arbitrary field polarization. 

At high energies, $b_0\gg1$, we enter a different regime, where the dominant contribution to trident comes from the one-step part. Interestingly, the dominant contribution in this regime comes from a one-step term ($\mathbb{P}_{\rm dir}^{22\to1}$) that can be obtained with the gluing method\footnote{There is another one-step term ($\mathbb{P}_{\rm dir}^{11}$) which is also visible in the high-$b_0$ plots. This term happens to be quite small compared to $\mathbb{P}_{\rm dir}^{22\to1}$, but it does remain also for larger $b_0$. Unlike $\mathbb{P}_{\rm dir}^{22}$, $\mathbb{P}_{\rm dir}^{11}$ stays quite close to its LCF approximation.}; it is obtained in the same way as the two-step but by including the second step-function combination in~\eqref{StepsForStepsSep} rather than the first. So, at least for trident we can greatly increase the parameter region where the gluing approximation works by including all of $\theta(\theta_{42})\theta(\theta_{31})$ in~\eqref{StepsForStepsSep} rather than just the $\theta(\sigma_{43}-\sigma_{21})$ term.
If we compare plots in Fig.~\ref{sectionsFig2} with equal $b_0$ (not $\chi$), that is diagonally in Fig.~\ref{sectionsFig2}, we see that $b_0$ determines the ratio of the one-step ($\sim\mathbb{P}_{\rm dir}^{22\to1}$) and the two-step peaks, and that the width of this peak decreases with $a_0$. 

A comparison of the distributions in Fig.~\ref{sectionsFig1} and Fig.~\ref{sectionsFig2} with the corresponding ones in~\cite{Dinu:2019wdw} shows that, at larger $b_0$ the distributions look very similar for linear and circular polarization (apart from the spikes in the linear case coming from the saddle points/fast variations in the effective mass), while for low $b_0$ they are quite different. To understand this, note that we have defined $a_0$ in the circular case~\eqref{circularPolDefinition} such that the integral of ${\bf a}'^2(\phi)$ over one period is the same as in the linear case (neglecting the variation of the pulse envelope). This means that, given a value of $a_0$, the maximum field strength is lower in the circular case.
At low $b_0$ this means a stronger exponential suppression for circular polarization, and a higher discrepancy from LCF. At large $b_0$, on the other hand, the formation length is large and the average intensity is more relevant, which explains why the linear and circular cases are more similar. 
Note also that the shape of the dominant one-step term ($\mathbb{P}_{\rm dir}^{22\to1}$) in Fig.~\ref{sectionsFig1} and Fig.~\ref{sectionsFig2} looks remarkably similar to the linear case, even for low $b_0$.

\begin{figure}
\includegraphics[width=\linewidth]{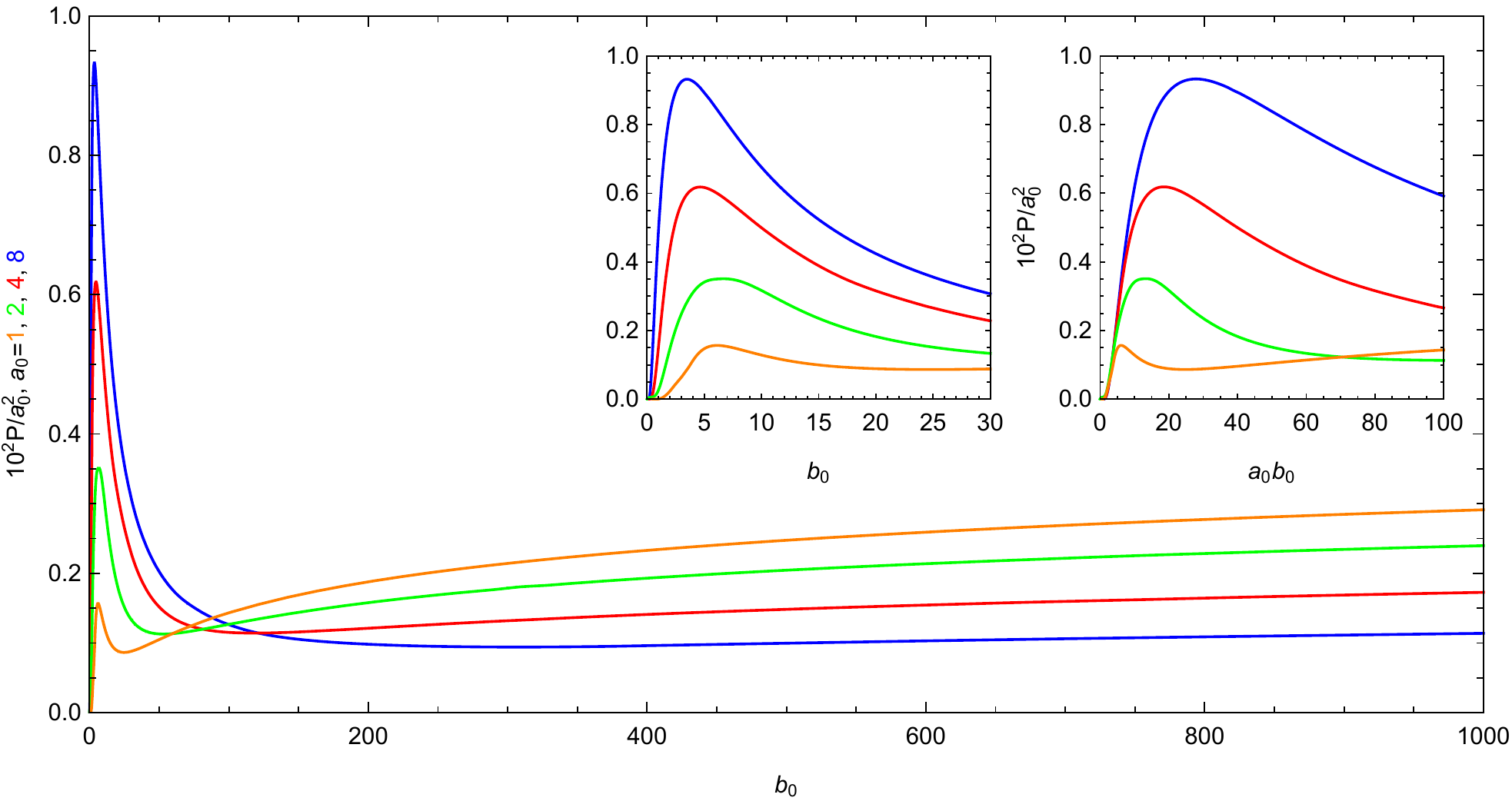}
\caption{Integrated probability as function of energy parameter $b_0$.}
\label{totalFig}
\end{figure}

For completeness, in Fig.~\ref{3DfigLowb0} and~\ref{3DfigHighb0} we show the full 3D spectrum for different values of $a_0$ and $\chi$. The overall shapes resemble those for linear polarization in~\cite{Dinu:2019wdw}, except that the circular ones are much smoother. This is what one can expect, e.g. because the ($\phi$-dependent) effective mass varies more slowly in a circularly polarized field. 
At low $\chi$ the distribution is significant in a limited region at the center of the $s_i$ triangle. As $\chi$ increases, the peak grows and starts to fill a large part of the triangle. As we reach high values of $b_0$, the distribution starts again to concentrate, this time into two sharp peaks located close to the $s_1=1$ and $s_2=1$ corners. These peaks become progressively taller, but also thinner, so the total probability will grow, but only slowly, as seen in Fig.~\ref{totalFig}.
Note also the peak close to $s_3=1$ for $a_0=1$ and $\chi=8$, where the positron takes most of the initial longitudinal momentum (compared to the peaks at $s_1=1$ and $s_2=1$ where one of the electrons takes most of the momentum).

An important question is how the probability scales with $a_0$ and what $b_0$ values allow us to maximize it. Scaling the integrated probability by $1/a_0^2$ allows us to compare the probabilities for $a_0=1,2,4,8$ in Fig.~\ref{totalFig}. It shows that the probability reaches a maximum at a finite $b_0$ (cf. similar plots for nonlinear Breit-Wheeler pair production in~\cite{Dinu:2013gaa}). 
Even with the LCF-inspired scaling, $\mathbb{P}/a_0^2$, the height increases with a0. The position of the peak can be seen in the two insets. The $a_0=8$ peak corresponds to the smallest $b_0$ value, but to the largest $\chi$. Increasing $b_0$ above this value will not help in achieving a greater or comparable probability, unless we increase $b_0$ to an extremely high value.
However, for low $a_0$, or a shorter pulse, one can beat the peak by going into the slow-growth region of higher $b_0$. This can be seen explicitly in Fig.~\ref{totalFig} for $a_0=1$.
If the pulse energy is small, at even moderately high $b_0$ values, the growth in the one-step will be able to provide us with a much larger probability than the peak we find in the two-step term.

\section{Saddle-point approximation}\label{saddlePointSection}

\begin{figure}
\includegraphics[width=1\linewidth]{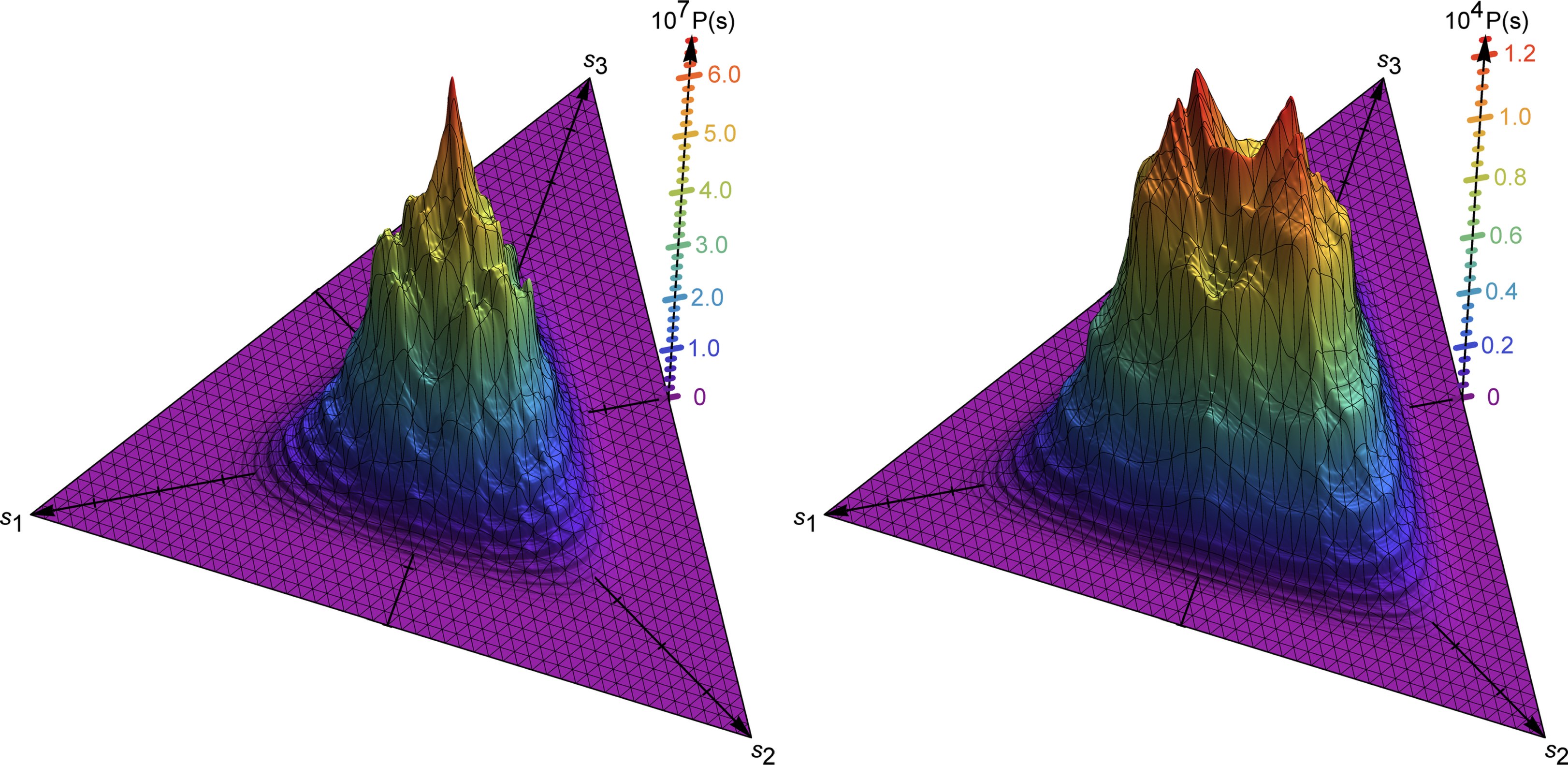}
\caption{The saddle point approximation of the spectrum for $a_0=1$, $\mathcal{T}=80$ and $\chi=0.5,1$.}
\label{saddleSpec1}
\end{figure}

In this section we will present a method that can be used for a first approximation of the higher-order processes considered above.
In~\cite{Dinu:2018efz} we used a saddle-point approach to obtain an approximation for $\chi<1$ of the spectrum for Compton scattering with a remarkably good agreement with the exact numerical result, including small and fast oscillations. In this section we use the same method for trident.
Saddle-point/semi-classical methods are of course often used for various strong-field processes, see~\cite{Meuren:2015mra,Nousch:2015pja} for two recent studies of Breit-Wheeler pair production. 
For the field we consider here
\be
a(\phi)=a_0\sin(\phi)e^{-(\phi/\mathcal{T})^2} \;,
\ee
we can perform the integrals in $\Theta_{ij}$ and $\Delta$ analytically in terms of error functions. 
We consider linear polarization which is simpler in this approach and also leads to a richer spectrum. 
The saddle points are determined by these two equations
\be
\frac{\partial\Theta_{21}}{\partial\sigma_{21}}=\frac{\partial\Theta_{21}}{\partial\theta_{21}}=0
\ee 
and similar for $\sigma_{43}$ and $\theta_{43}$. Note that these are exactly the same saddle-point equations as in~\cite{Dinu:2018efz} for Compton scattering. So, we can reuse the saddle points we already have, and we refer to~\cite{Dinu:2018efz} for the details on how to obtain them.
For a field with many oscillations there is a large number of saddle points to include, which all lie in the complex plane. For one-dimensional integrals one can deform the original integration contour to a sum over steepest descent contours that go through some saddle points, but not necessarily all saddle points. For multidimensional integrals like the ones we have here, it is much more nontrivial to construct the higher-dimensional version of steepest descent contours and it is in general a nontrivial question which saddle points one should actually include. (These questions are also considered in Monte-Carlo/Lefschetz thimbles approaches to e.g. the sign problem in QCD~\cite{Cristoforetti:2012su}.) We should of course not include the saddle points that give exponentially large contributions, but by including the other saddle points we have found a good agreement with the exact numerical result.
For a pulsed oscillating field we in general have to obtain these saddle points numerically. To do so we need starting points. As explained in~\cite{Dinu:2018efz} we obtain the saddle points by using the corresponding ones for a monochromatic field (which are easier to find) as starting points. The saddle points move continuously through the complex plane as we decrease the pulse length $\mathcal{T}$ (or change $a_0$), and in some cases it can be useful to consider a couple of intermediate values of $\mathcal{T}$ between the monochromatic case $\mathcal{T}=\infty$ and the actual value of $\mathcal{T}$.
As explained in~\cite{Dinu:2018efz}, each pair of $\sigma$ and $\theta$ have a set of saddle points which are characterized by two integers, $n$ and $m$, where increasing $n$ and $m$ correspond, respectively, to increasing $\text{Re}\,\sigma$ and $\text{Re}\,\theta$ (see~\cite{Dinu:2018efz} for the exact definition of $n$ and $m$). 
There are two sets of saddle points. Here the dominant contribution comes from the ones that are continuously connected to 
\be
\{\sigma,\theta\}=\left\{n\pi,2i\text{arcsinh}\left[\frac{1}{a_0}\right]+2m\pi\right\}
\ee
in the monochromatic limit $\mathcal{T}\to\infty$.

A new aspect compared to the first order processes is that now we have step functions for the $\phi$ integrals, which lead to restrictions on which saddle points to include.
For $\mathbb{P}_{\rm dir}^{22\to2}$ we have $\theta(\sigma_{43}-\sigma_{21})$ and therefore we should only include saddle points with $n_{43}\geq n_{21}$. For $n_{43}=n_{21}$ the step function removes one half of a Gaussian integral, which gives an overall factor of $1/2$ compared to the cases with $n_{43}>n_{21}$. For $\mathbb{P}_{\rm dir}^{22}$ we have a more complicated step function, $\theta\left(\sigma_{43}-\sigma_{21}-\frac{|\theta_{43}-\theta_{21}|}{2}\right)$. For saddle points with $\sigma_{43}-\sigma_{21}=\frac{|\theta_{43}-\theta_{21}|}{2}>0$ one can again diagonalize the quadratic fluctuations around the saddle point such that the step functions simply remove one half of one of the Gaussian integrals. For saddle points with $\sigma_{43}-\sigma_{21}=\frac{|\theta_{43}-\theta_{21}|}{2}=0$ the step function restricts two Gaussian integrals and we have integrals on the form 
\be
\begin{split}
&\int_0^\infty\ud x\ud y e^{-ax^2-by^2-cxy} \\
&=\frac{1}{2\pi}\text{arccos}\frac{c}{2\sqrt{ab}}\int_{-\infty}^\infty\ud x\ud y e^{-ax^2-by^2-cxy} \;,
\end{split}
\ee  
where the factor in front of the integral in the second line gives the relative factor compared to the case without a step function. 

Note that to find the saddle points numerically, we only have to specify $a_0$ and $\mathcal{T}$, so we have
\be
\begin{split}
\mathbb{P}(s)=&\alpha^2\frac{1+[1-\kappa_{01}][1+\kappa_{23}]}{q_1^2r_1r_2}\!\sum_{n_{21},m_{21},n_{43},m_{43}}\!\text{``prefactor''} \\
&\exp\left\{\frac{i}{2b_0}[r_1\Theta_{21}+r_2\Theta_{43}]\right\}+(1\leftrightarrow2) \;,
\end{split}
\ee     
where ``prefactor'' and $\Theta_{ij}$ depend on $a_0$ and $\mathcal{T}$ via the numerically obtained saddle points (the Gaussian integrals, obtained from expanding to second order around the saddle points, are performed analytically and then evaluated by inserting the numerical saddle points), but not on the momenta $s_i$ or $b_0$. 

The saddle-point approximation can be expected to be good when $\chi$ is sufficiently small.
As can be seen by comparing Fig.~\ref{saddleSpec1} with the corresponding exact results in~\cite{Dinu:2019wdw}, this saddle-point approximation captures many of the features of the exact result even for $\chi=0.5$, which is not particularly small. The most easily seen difference is the slightly higher peak in the saddle-point approximation. For $\chi=1$ we start to see a bit larger differences also in the shape of the spectrum, but the saddle-point result still gives a good first approximation.  
Producing the data for the saddle-point approximation in Fig.~\ref{saddleSpec1} is of course much faster than obtaining the exact results. So, the saddle-point method can be used to quickly test and find interesting parameter values, for which one can then produce more accurate results with an exact integration, e.g. with the methods described in~\cite{Dinu:2019wdw}.

\section{Conclusions}

For sufficiently large $a_0$ or sufficiently long pulse, one can expect the dominant contribution to trident to come from the incoherent product of nonlinear Compton scattering and Breit-Wheeler pair production. The nontrivial problem is how to treat the polarization of the intermediate photon. For constant-crossed fields it was already known~\cite{Ritus:1972nf,Baier,King:2013osa} that the two-step can be obtained by summing the incoherent product over two suitable polarizations vectors. In~\cite{Dinu:2017uoj} we showed that this simple sum can be generalized to inhomogeneous fields for $a_0\sim1$ (the two-step in LCF/constant-crossed fields corresponds to the leading order term in an expansion in $1/a_0\ll1$). In this paper we have studied how to generalize this to general (e.g. circular) field polarization. This turns out to be nontrivial. We have managed to find such a generalization, and it is not simply a (single) sum over the polarization of the intermediate photon. This gluing generalization involves a 3D unit (Stokes) vector describing the photon polarization. In~\cite{Dinu:2018efz} we provided a similar gluing generalization to double nonlinear Compton scattering, where the intermediate particle is an electron instead of a photon. In that case the spin of the intermediate electron enters via another 3D unit vector (and the two-step is again not simply a sum over two independent spins). Interestingly, although the 3D unit vectors for the photon polarization and electron spin are different objects and transforms differently, they enter the construction of gluing estimate in basically the same way. Now, if this gluing approach only worked for the second-order processes, trident and double Compton, it would perhaps not have been so useful, because we have calculated all contributions to the probability independently of this gluing approximation and so we anyway know what the two-step is and what the gluing approach has to give. However, in this paper we have generalized to higher orders and showed that the $N$-step part of N'th-order processes with $N>2$ (which can be obtained in a spin/polarization basis independent way with Dirac traces of $\slashed{p}+1$ etc.) can be obtained with the same gluing approach.
Thus we have provided basic building blocks with which to construct estimates of general higher-order processes.

\acknowledgments
V.~Dinu has been supported by a grant from Romanian National Authority for Scientific Research, CNCS-UEFISCDI, Project No. PN-III-P4-IDPCCF-2016-0164, within PNCDI III. 
G.~Torgrimsson was supported by the Alexander von Humboldt foundation.

\appendix

\end{document}